\documentclass[
aps, % Option for submissions to the American Physical Society journals.
pra, % Option for submissions to the Physical Review A journal.
reprint, % Option for the style with two columns.
floatfix, % Option to invoke emergency float processing.
footinbib, % Option to put footnote text at the bottom of a page.
longbibliography, % Option for the bibliography style with titles of the literature.
]{revtex4-1}

\usepackage{amsmath} % Package for sophisticated mathematical typesetting.
\usepackage{amsfonts} % Extended set of mathematical fonts.
\usepackage{amssymb} % Extended set of mathematical symbols.
\usepackage{amsthm} % Package for typesetting of theorems.
\usepackage{mathtools} % Additional tools to use with amsmath package.
\usepackage{braket} % Package for implementation of the bra and ket symbols from the quantum mechanics.
\usepackage{graphicx} % Package allowing precise control over the imported images.
\usepackage{enumerate} % Package for enumerate environment.
\usepackage{enumitem} % Package for modifying enumerate environment.
\usepackage{tabularx} % Package for creation of a paragraph-like column expanding to fill the environment.
\usepackage[breaklinks]{hyperref} % Package for implementation of the hyperlinks.
\usepackage[dvipsnames]{xcolor} % Package used to set the font color, text background, or page background.

\theoremstyle{definition}

\newtheorem{definition}{Definition of the $\epsilon$-security}
\newtheorem*{definition*}{Definition of the $\epsilon$-security}

\newtheorem{theorem}[definition]{Theorem}
\newtheorem*{theorem*}{Theorem}

\newtheorem{lemma}[definition]{Lemma}
\newtheorem*{lemma*}{Lemma}

\newtheorem{property}[definition]{Property}
\newtheorem*{property*}{Property}

\newcommand*{\ancillas}{\mathrm{ancillas}}

\newcommand{\overbar}[1]{\mkern 1.5mu\overline{\mkern-1.5mu#1\mkern-1.5mu}\mkern 1.5mu}

\begin{document}

\title{Secure multi-party quantum computation protocol for quantum circuits: \\
the exploitation of triply-even quantum error-correcting codes}

\author{Petr A. Mishchenko}
\email{petr.mishchenko.us@hco.ntt.co.jp}

\author{Keita Xagawa}
\email{keita.xagawa.zv@hco.ntt.co.jp}

\affiliation{NTT Social Informatics Laboratories, Tokyo 180-8585, Japan}

\date{\today}

% /////////////////////////////////////////////////////////////////////////////////////////////////////////////////////////////////////////
% Abstract (begin)
% /////////////////////////////////////////////////////////////////////////////////////////////////////////////////////////////////////////

\begin{abstract}
Secure multi-party quantum computation (MPQC) protocol is a cryptographic primitive allowing error-free distributed quantum computation to
a group of $n$ mutually distrustful quantum nodes even when some quantum nodes disobey the instructions of the protocol. Here we suggest a
modified MPQC protocol that adopts unconventional quantum error-correcting codes and as a consequence reduces the number of qubits required
for the protocol execution. In particular, the replacement of the self-dual Calderbank-Shor-Steane quantum error-correcting codes with
triply-even ones permits us to avoid the previously indispensable but resource-intensive procedure of the ``magic'' state verification.
Besides, since every extra qubit reduces the credibility of physical devices, our suggestion makes the MPQC protocol more accessible for
the near-future technology by reducing the number of necessary qubits per quantum node from $n^2 + \Theta(r)n$, where $r$ is the security
parameter, to $n^2 + 3n$.
\end{abstract}

% /////////////////////////////////////////////////////////////////////////////////////////////////////////////////////////////////////////
% Abstract (begin)
% /////////////////////////////////////////////////////////////////////////////////////////////////////////////////////////////////////////

\maketitle

% /////////////////////////////////////////////////////////////////////////////////////////////////////////////////////////////////////////
\section{
  \label{sec:introduction}
  Introduction
}
% /////////////////////////////////////////////////////////////////////////////////////////////////////////////////////////////////////////

As a well-established and widely used tool secure multi-party classical computation (MPCC) protocol allows $n$ classical nodes to jointly
compute some publicly known function $y = f(x^1, \ldots, x^n)$ on their private inputs $x^1, \ldots, x^n$ in a distributed
manner~\cite{yao_sfcs_23_160-164_1982}. During the execution of the MPCC protocol cheating classical nodes, following the instructions of
the protocol not honestly, cannot affect the output of the computation $y$ beyond choosing their inputs as well as cannot obtain any
information on the inputs of the honest classical nodes beyond what they can infer from the output of the computation $y$. Since the MPCC
protocol allows for distributed computation of any function $f$ it becomes a powerful cryptographic primitive with many practical
applications, e.g., secure electronic auction, secure electronic voting, and secure machine learning~\cite{evans_kolesnikov_rosulek}.

\begin{figure}[htb]
  \begin{center}
% !!!!!!!!!!!!!!!!!!!!!!!!!!!!!!!!!!!!!!!!!!!!!!!!!!!!!!!!!!!!!!!!!!!!!!!!!!!!!!!!!!!!!!!!!!!!!!!!!!!!!!!!!!!!!!!!!!!!!!!!!!!!!!!!!!!!!!!!!
    \includegraphics[width=\columnwidth]{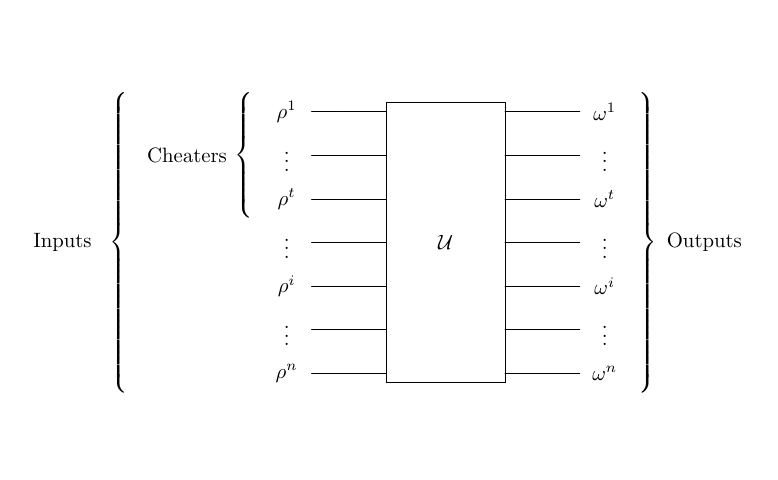}
% !!!!!!!!!!!!!!!!!!!!!!!!!!!!!!!!!!!!!!!!!!!!!!!!!!!!!!!!!!!!!!!!!!!!!!!!!!!!!!!!!!!!!!!!!!!!!!!!!!!!!!!!!!!!!!!!!!!!!!!!!!!!!!!!!!!!!!!!!
    \caption{
      Schematic picture of the MPQC protocol for the quantum circuit model of computation. At the beginning of the MPQC protocol, each
      quantum node $i$ provides an input quantum state $\rho^i$. Then, $n$ quantum nodes jointly perform quantum circuit $\mathcal{U}$ with
      $n$ inputs and $n$ outputs. At the end of the MPQC protocol, each quantum node $i$ receives an output quantum state $\omega^i$. The
      purpose of the MPQC protocol is to implement quantum circuit $\mathcal{U}$ in such a way that requirements of \textit{correctness},
      \textit{soundness}, and \textit{privacy} are satisfied even in the presence of $t < \frac{n}{4}$ cheating quantum nodes.
    }
% !!!!!!!!!!!!!!!!!!!!!!!!!!!!!!!!!!!!!!!!!!!!!!!!!!!!!!!!!!!!!!!!!!!!!!!!!!!!!!!!!!!!!!!!!!!!!!!!!!!!!!!!!!!!!!!!!!!!!!!!!!!!!!!!!!!!!!!!!
    \label{fig:mpqc_sketch}
% !!!!!!!!!!!!!!!!!!!!!!!!!!!!!!!!!!!!!!!!!!!!!!!!!!!!!!!!!!!!!!!!!!!!!!!!!!!!!!!!!!!!!!!!!!!!!!!!!!!!!!!!!!!!!!!!!!!!!!!!!!!!!!!!!!!!!!!!!
  \end{center}
\end{figure}

Later developed more powerful quantum approach, secure multi-party quantum computation (MPQC) protocol, allows $n$ quantum nodes to jointly
compute some publicly known quantum circuit $\mathcal{U}(\rho^1, \ldots, \rho^n)$ on their private inputs
$\rho^1, \ldots, \rho^n$~\cite{crepeau_stoc_02_643-652_2002}. In more detail, MPQC protocol can be described as a cryptographic primitive
where each quantum node $i$ inputs some quantum state $\rho^i$, and then $n$ quantum nodes jointly perform arbitrary quantum circuit
$\mathcal{U}$ with $n$ inputs and $n$ outputs. Finally, each quantum node $i$ obtains some output quantum state $\omega^i$. See the
schematic representation of the MPQC protocol in Fig.~\ref{fig:mpqc_sketch}. Akin to its classical counterpart, the MPQC protocol satisfies
the following informal requirements even in the presence of cheating quantum nodes:
\begin{itemize}
  \item \textbf{Correctness and Soundness:} Cheating quantum nodes cannot affect the outcome of the MPQC protocol beyond choosing their
  inputs.
  \item \textbf{Privacy:} Cheating quantum nodes can learn nothing about the private inputs and outputs of the honest quantum nodes beyond
  what they can infer from the output of the computation.
\end{itemize}

Currently existing MPQC protocols developed for the quantum circuit model of computation can be divided into two types:
information-theoretically secure ones~\cite{smith_arxiv:quant-ph/0111030,crepeau_stoc_02_643-652_2002,ben-or_focs_06_249-260_2006,
lipinska_pra_102_022405_2020,goyal_cryptoeprint:2022/1583}, meaning that there are no assumptions on the computational power of the
cheating quantum nodes, and computationally secure ones~\cite{dupuis_crypto_6223_685-706_2010,dupuis_crypto_7417_794-811_2012,
dulek_eurocrypt_12107_729-758_2020,alon_crypto_12825_436-466_2021,bartusek_crypto_12825_406-435_2021,huang_cryptoeprint:2022/1517}. The
former type of the MPQC protocols is based on a technique of quantum error correction, which limits the maximum number of cheating quantum
nodes to $t < \frac{n}{4}$, i.e., a constraint inherent to the Knill-Laflamme bound or the so called quantum Singleton
bound~\cite{knill_pra_55_900_1997}, while the latter type of the MPQC protocols is based on a technique of quantum authentication codes and
can tolerate $t < n$ cheating quantum nodes.

% -----------------------------------------------------------------------------------------------------------------------------------------
% -----------------------------------------------------------------------------------------------------------------------------------------

In this paper, we consider the information-theoretically secure MPQC protocol which is based on a technique of quantum error correction. As
a matter of fact, the technique of quantum error correction is tightly related to the concept of quantum secret
sharing~\cite{cleve_prl_83_648_1999}, the verifiable version of which was first suggested in Ref.~\cite{crepeau_stoc_02_643-652_2002}, and
became a prevalent tool for constructing the information-theoretically secure MPQC protocols. In particular, following the approach taken
in Refs.~\cite{lipinska_pra_102_022405_2020,lipinska_arxiv:2004.10486v2} we utilize the verifiable hybrid secret sharing (VHSS) protocol
suggested in Ref.~\cite{lipinska_pra_101_032332_2020}, i.e., a modified version of the original verifiable quantum secret sharing protocol
presented in Ref.~\cite{crepeau_stoc_02_643-652_2002}. The VHSS protocol is rather versatile and works for any type of the
Calderbank-Shor-Steane (CSS) quantum error correcting codes (QECCs)~\cite{steane_prsla_452_2551-2577_1996,calderbank_pra_54_1098_1996}.
Therefore, at the beginning of the MPQC protocol, all the quantum nodes should agree on some CSS QECC with which they will remain until the
end of the MPQC protocol.

At the highest level of abstraction, the MPQC protocol built upon a technique of quantum error correction and associated with it verifiable
quantum secret sharing is executed in the following way. First of all, each quantum node $i$ encodes and shares his single-qubit input
quantum state $\rho^i$. In such a way, quantum nodes create a global logical quantum state shared among all the $n$ quantum nodes, and as a
consequence, each quantum node $i$ holds a part of the global logical quantum state which we call a \textit{share}. Next, quantum nodes
jointly verify the encoding of each single-qubit input quantum state $\rho^i$ using the VHSS protocol~\cite{lipinska_pra_101_032332_2020}.
Then, quantum nodes locally perform quantum operations on their \textit{shares} of the global logical quantum state to evaluate the logical
version of the quantum circuit $\mathcal{U}$. Finally, each quantum node $i$ collects all the \textit{shares} corresponding to his output
from the other quantum nodes and by decoding these \textit{shares} reconstructs his single-qubit output quantum state $\omega^i$. Note that
at this level of abstraction, our MPQC protocol follows the procedure of the previously suggested MPQC protocol in
Refs.~\cite{lipinska_pra_102_022405_2020,lipinska_arxiv:2004.10486v2}.

To implement the universal quantum computation (UQC) in the above MPQC protocol, a particular universal set of quantum gates need to be
chosen. In addition, these quantum gates need to be transversal for a CSS QECC that is chosen at the beginning of the MPQC protocol.
Specifically, this means that the application of local quantum operations to each \textit{share} should yield a meaningful logical
operation on the global logical quantum state. However, it is known to be impossible to implement an entire universal set of quantum gates
transversally not only for the CSS QECCs but for any QECC~\cite{eastin_prl_102_110502_2009}. Actually, the solution to this problem lies in
the extension of the transversal set of quantum gates with a non-transversal quantum gate, which can be implemented for example by the gate
teleportation technique, i.e., with the help of transversal quantum gates, local measurements, classical communication, and ancillary
quantum state~\cite{gottesman_nature_402_390-393_1999}.

In particular, the MPQC protocol suggested in Refs.~\cite{lipinska_pra_102_022405_2020,lipinska_arxiv:2004.10486v2} is based on a sub-class
of CSS QECCs, i.e., self-dual CSS QECCs~\cite{steane_prsla_452_2551-2577_1996,calderbank_pra_54_1098_1996} where the universal set of
quantum gates is chosen to be the Clifford gates ($H$ gate, $P$ gate, and $\mathrm{C}\text{-}X$ gate) in combination with the $T$
gate~\cite{nebe_dcc_24_99-122_2001}, see Appendix~\ref{sec:quantum_gates_definitions} for definitions of the quantum gates. Self-dual CSS
QECCs allow trivial implementation of the transversal Clifford gates but require additional techniques for implementation of the
non-transversal $T$ gate. In case of the MPQC protocol originally suggested in Ref.~\cite{lipinska_pra_102_022405_2020} and later
significantly reconsidered in Ref.~\cite{lipinska_arxiv:2004.10486v2} one requires two additional techniques: the gate teleportation
technique and the verification of the ``magic'' state technique, the latter of which is implemented by a statistical testing of the
randomly selected ``magic'' states with their subsequent distillation~\footnote{In the original version of the MPQC protocol the
verification of the ``magic'' state technique was implemented by the protocol called ``verification of the Clifford stabilized states''
(VCSS) which contained potential problems coming from the engagement of a non-transversal $\mathrm{C}\text{-}XP^\dag$ gate, see
Appendix~\ref{sec:vcss_summary} and Appendix~\ref{sec:vcss_critics}.}. Indeed, these additional techniques require an extra workspace for
the implementation.

In this paper, we suggest an MPQC protocol constructed on the basis of triply-even CSS QECCs~\cite{betsumiya_jlms_86_1_1-16_2012,
knill_arxiv:quant-ph/9610011}, which constitute an another sub-class of general CSS QECCs~\cite{steane_prsla_452_2551-2577_1996,
calderbank_pra_54_1098_1996}. In case of triply-even CSS QECCs, we decide on another universal set of quantum gates, i.e., $X$ gate, $Z$
gate, $T$ gate, $\mathrm{C}\text{-}X$ gate, and $H$ gate~\cite{chamberland_qst_2_035008_2017}, among which, triply-even CSS QECCs allow
transversal implementation of $X$ gate, $Z$ gate, $T$ gate, and $\mathrm{C}\text{-}X$ gate while do not allow transversal implementation of
the $H$ gate~\cite{knill_arxiv:quant-ph/9610011}. Therefore, in our MPQC protocol, a non-transversal $H$ gate is implemented by the gate
teleportation technique, which has a lot of similarities with the implementation of the $T$ gate in
Refs.~\cite{lipinska_pra_102_022405_2020,lipinska_arxiv:2004.10486v2}. Nonetheless, the implementation of the non-transversal $H$ gate by
the gate teleportation technique does not require verification of the ancillary logical ``magic'' state, i.e., whether it is certainly
the logical ``magic'' state or not, see Ref.~\cite{lipinska_arxiv:2004.10486v2} and Appendix~\ref{sec:magic_state_verification_protocol},
since as an ancillary quantum state we need the logical ``plus'' state, i.e., a logical version of the single-qubit quantum state
$\ket{+} = \frac{1}{\sqrt{2}}(\ket{0} + \ket{1})$. Unlike the case of the logical ``magic'' state verification, the logical ``plus'' state
can be easily verified by using the VHSS protocol only. Therefore, by avoiding the verification of the ancillary logical ``magic'' state we
can reduce the workspace required for the implementation of the MPQC protocol from $n^2 + \Theta(r)n$ qubits in case of the previous
suggestion, see Ref.~\cite{lipinska_arxiv:2004.10486v2}, to $n^2 + 3n$ qubits in our case, where $n$ is the number of quantum nodes
participating in the MPQC protocol and $r$ is the security parameter.

The paper is organized as follows. In Sec.~\ref{sec:mpqc_summary}, we briefly overview our MPQC protocol. Then, in
Sec.~\ref{sec:assumptions_and_definitions}, we declare our assumptions and definitions necessary for the construction of the MPQC protocol.
We describe our assumptions on communication channels in Sec.~\ref{subsec:communication_channels}, and our assumptions on the properties of
the adversary in Sec.~\ref{subsec:adversary}. In Sec.~\ref{subsec:css_qeccs}, we define properties common to any type of CSS QECCs, and in
Sec.~\ref{subsec:triply-even_css_qeccs}, we discuss a sub-class of general CSS QECCs called triply-even CSS QECCs. Subsequently, in
Sec.~\ref{sec:mpqc_subroutines}, we outline all the subroutines involved in the MPQC protocol: the VHSS protocol in
Sec~\ref{subsec:vhss_protocol} and the gate teleportation protocol in Sec.~\ref{subsec:gate_teleportation_protocol}. After that, in
Sec.~\ref{sec:mpqc_protocol}, we present a detailed outline of our MPQC protocol. Next, in Sec.~\ref{sec:security_mpqc_protocol}, we prove
the security of our MPQC protocol. In particular, we begin with stating the security framework as well as the security definition of our
MPQC protocol in Sec.~\ref{subsec:security_statements}, and then we find that the security proof of our MPQC protocol should be identical
to the previously suggested MPQC protocol in Sec.~\ref{subsec:security_proof}. Moreover, to be self-contained, we briefly present the
security proof of our MPQC protocol in Secs.~\ref{subsubsec:ideal_protocol} and \ref{subsubsec:real_protocol}. Finally,
Sec.~\ref{sec:summary} is devoted to the summary.

% /////////////////////////////////////////////////////////////////////////////////////////////////////////////////////////////////////////
\section{
  \label{sec:mpqc_summary}
  Summary of the MPQC protocol
}
% /////////////////////////////////////////////////////////////////////////////////////////////////////////////////////////////////////////

% -----------------------------------------------------------------------------------------------------------------------------------------
\begin{table}[htb]
% !!!!!!!!!!!!!!!!!!!!!!!!!!!!!!!!!!!!!!!!!!!!!!!!!!!!!!!!!!!!!!!!!!!!!!!!!!!!!!!!!!!!!!!!!!!!!!!!!!!!!!!!!!!!!!!!!!!!!!!!!!!!!!!!!!!!!!!!!
  \caption{
    Summary of the MPQC protocol.
  }
% !!!!!!!!!!!!!!!!!!!!!!!!!!!!!!!!!!!!!!!!!!!!!!!!!!!!!!!!!!!!!!!!!!!!!!!!!!!!!!!!!!!!!!!!!!!!!!!!!!!!!!!!!!!!!!!!!!!!!!!!!!!!!!!!!!!!!!!!!
  \begin{ruledtabular}
    \begin{tabularx}{\linewidth}{X}
% !!!!!!!!!!!!!!!!!!!!!!!!!!!!!!!!!!!!!!!!!!!!!!!!!!!!!!!!!!!!!!!!!!!!!!!!!!!!!!!!!!!!!!!!!!!!!!!!!!!!!!!!!!!!!!!!!!!!!!!!!!!!!!!!!!!!!!!!!
      \\
      \textbf{Input:} Single-qubit quantum state $\rho^i$ from each quantum node $i$, agreement on a particular $\mathcal{C}_\mathrm{TE}$,
      and a particular $\mathcal{U}$. \\ \\
      \textbf{Output:} In case of success, single-qubit quantum state $\omega^i$ in the possession of each quantum node $i$. In case of
      failure, i.e., excess in the number of cheating quantum nodes, the MPQC protocol is aborted at the end of the computation.
      \\
% !!!!!!!!!!!!!!!!!!!!!!!!!!!!!!!!!!!!!!!!!!!!!!!!!!!!!!!!!!!!!!!!!!!!!!!!!!!!!!!!!!!!!!!!!!!!!!!!!!!!!!!!!!!!!!!!!!!!!!!!!!!!!!!!!!!!!!!!!
      \begin{enumerate}
        \item \textbf{Sharing:} By encoding and sharing each of the inputs $\rho^1, \ldots, \rho^n$ \textit{twice}, quantum nodes create a
        global logical quantum state $\bar{\bar{\mathcal{P}}}$ where each quantum node $i$ holds a share $\bar{\bar{\mathcal{P}}}_i$. For
        the details see Sec.~\ref{subsec:vhss_protocol}.
        \item \textbf{Verification:} All the quantum nodes jointly verify the encoding of each input $\rho^i$ with the help of the VHSS
        protocol to check whether each quantum node $i$ is honest or not. For the details see Sec.~\ref{subsec:vhss_protocol}.
        \item \textbf{Computation:} Depending on whether the quantum gate appearing in $\mathcal{U}$ can be implemented transversally or
        not, or whether the implementation of the $\mathcal{U}$ requires an ancillary quantum state, quantum nodes behave in the following
        three ways:
        \begin{enumerate}
          \item In case of the transversal quantum gates, i.e., $X$ gate, $Z$ gate, $T$ gate, or $\mathrm{C}\text{-}X$ gate, each quantum
          node $i$ locally applies corresponding quantum operations to his share $\bar{\bar{\mathcal{P}}}_i$.
          \item In case of the non-transversal $H^i$ gate applied to the quantum wire $i$ of the $\mathcal{U}$, quantum nodes jointly
          prepare verified by the VHSS protocol ancillary logical quantum state $\bar{\bar{\ket{+}}}^i$ created from the single-qubit
          quantum state $\ket{+}^i$, and then perform the gate teleportation technique. For the details see
          Sec.~\ref{subsec:gate_teleportation_protocol}.
          \item In case the implementation of the $\mathcal{U}$ requires an ancillary quantum state, quantum nodes jointly prepare verified
          by the VHSS protocol ancillary logical quantum state $\bar{\bar{\ket{0}}}^i$ created from the single-qubit quantum state
          $\ket{0}^i$.
        \end{enumerate}
        \item \textbf{Reconstruction:} Each quantum node $i$ collects all the single-qubit quantum states corresponding to his output and
        by decoding in such a way obtained output logical quantum state $\bar{\bar{\Omega}}^i$ \textit{twice}, reconstructs his output
        $\omega^i$. For the details see Sec.~\ref{subsec:vhss_protocol}.
      \end{enumerate}
% !!!!!!!!!!!!!!!!!!!!!!!!!!!!!!!!!!!!!!!!!!!!!!!!!!!!!!!!!!!!!!!!!!!!!!!!!!!!!!!!!!!!!!!!!!!!!!!!!!!!!!!!!!!!!!!!!!!!!!!!!!!!!!!!!!!!!!!!!
      \\
% !!!!!!!!!!!!!!!!!!!!!!!!!!!!!!!!!!!!!!!!!!!!!!!!!!!!!!!!!!!!!!!!!!!!!!!!!!!!!!!!!!!!!!!!!!!!!!!!!!!!!!!!!!!!!!!!!!!!!!!!!!!!!!!!!!!!!!!!!
    \end{tabularx}
  \end{ruledtabular}
% !!!!!!!!!!!!!!!!!!!!!!!!!!!!!!!!!!!!!!!!!!!!!!!!!!!!!!!!!!!!!!!!!!!!!!!!!!!!!!!!!!!!!!!!!!!!!!!!!!!!!!!!!!!!!!!!!!!!!!!!!!!!!!!!!!!!!!!!!
  \label{table:mpqc_outline}
% !!!!!!!!!!!!!!!!!!!!!!!!!!!!!!!!!!!!!!!!!!!!!!!!!!!!!!!!!!!!!!!!!!!!!!!!!!!!!!!!!!!!!!!!!!!!!!!!!!!!!!!!!!!!!!!!!!!!!!!!!!!!!!!!!!!!!!!!!
\end{table}
% -----------------------------------------------------------------------------------------------------------------------------------------

Here we briefly describe our MPQC protocol. In a similar manner to Refs.~\cite{crepeau_stoc_02_643-652_2002,smith_arxiv:quant-ph/0111030,
lipinska_pra_102_022405_2020,lipinska_arxiv:2004.10486v2}, our MPQC protocol is based on a technique of quantum error correction, or to be
more specific, on the concept of quantum secret sharing~\cite{cleve_prl_83_648_1999}, and in particular utilizes the VHSS protocol
suggested in Ref.~\cite{lipinska_pra_101_032332_2020} as its building block. The VHSS protocol works for any type of CSS QECCs encoding
single-qubit quantum states into $n$-qubit logical quantum states, see Sec.~\ref{subsec:vhss_protocol}. Therefore, in our MPQC protocol the
input quantum states $\rho^1, \ldots, \rho^n$ and the output quantum states $\omega^1, \ldots, \omega^n$ will indeed be single-qubit
quantum states. In particular, our construction of the MPQC protocol relies on a sub-class of general CSS
QECCs~\cite{steane_prsla_452_2551-2577_1996,calderbank_pra_54_1098_1996}, see Sec.~\ref{subsec:css_qeccs}, i.e., triply-even CSS QECCs
$\mathcal{C}_\mathrm{TE}$~\cite{betsumiya_jlms_86_1_1-16_2012,knill_arxiv:quant-ph/9610011}, see Sec.~\ref{subsec:triply-even_css_qeccs}.
In fact, triply-even CSS QECCs $\mathcal{C}_\mathrm{TE}$ allow transversal implementation of $X$ gate, $Z$ gate, $T$ gate, and
$\mathrm{C}\text{-}X$ gate while do not allow transversal implementation of the $H$ gate, and to implement the non-transversal $H$ gate we
utilize the gate teleportation technique as will be explained below, see Sec.~\ref{subsec:gate_teleportation_protocol}.

At the beginning of the MPQC protocol, quantum nodes should agree on a particular triply-even CSS QECC $\mathcal{C}_\mathrm{TE}$ described
above and then create global logical quantum state $\bar{\bar{\mathcal{P}}}$ by encoding and sharing each of the inputs
$\rho^1, \ldots, \rho^n$ \textit{twice}, see Fig.~\ref{fig:vhss_sharing}. Hereafter, the double bar always means that the quantum state is
encoded \textit{twice}. As a result, each quantum node $i$ holds a part of the global logical quantum state $\bar{\bar{\mathcal{P}}}$,
i.e., a share denoted as $\bar{\bar{\mathcal{P}}}_i$. Next, to check whether each quantum node $i$ is honest or not, quantum nodes jointly
verify the encoding of each input $\rho^i$ by using the VHSS protocol~\cite{lipinska_pra_101_032332_2020}, see
Sec.~\ref{subsec:vhss_protocol}. After that, quantum nodes jointly evaluate logical quantum circuit $\bar{\bar{\mathcal{U}}}$, i.e., a
\textit{twice} encoded version of the quantum circuit $\mathcal{U}$, see Sec.~\ref{sec:mpqc_protocol}. Here, in case of the transversal
quantum gates, each quantum node $i$ locally performs necessary quantum operations on his share $\bar{\bar{\mathcal{P}}}_i$. On the other
hand, in case of non-transversal quantum gates, quantum nodes jointly perform the gate teleportation technique, see
Sec.~\ref{subsec:gate_teleportation_protocol}. In addition, if the implementation of the logical quantum circuit $\bar{\bar{\mathcal{U}}}$
requires an ancillary quantum state, quantum nodes jointly create the ancillary logical quantum state $\bar{\bar{\ket{0}}}^i$ by encoding
and sharing a single-qubit quantum state $\ket{0}^i$ \textit{twice}. Finally, each quantum node $i$ collects all the single-qubit quantum
states corresponding to his output from the other quantum nodes and by decoding in such a way obtained output logical quantum state
$\bar{\bar{\Omega}}^i$ \textit{twice}, eventually reconstructs his output $\omega^i$, see Sec.~\ref{subsec:vhss_protocol}. Also, during the
execution of the MPQC protocol quantum nodes publicly record the positions of the cheating quantum nodes to decide whether to abort the
MPQC protocol or not. In particular, information on the positions of the cheating quantum nodes is updated each time the VHSS protocol or
the gate teleportation protocol is invoked.

In short, the gate teleportation technique implementing a non-transversal $H^i$ gate, where superscript $i$ means that the quantum gate is
applied to the quantum wire $i$ of the quantum circuit $\mathcal{U}$, is performed as follows. Suppose quantum nodes want to apply a
non-transversal $\bar{\bar{H}}^i$ gate, i.e., a logical version of the non-transversal $H^i$ gate, applied to the part of the global
logical quantum state $\bar{\bar{\mathcal{P}}}$ initially created from the single-qubit input quantum state $\rho^i$ and denoted as
$\bar{\bar{\mathcal{P}}}^i$. Quantum nodes jointly prepare verified by the VHSS protocol ancillary logical quantum state
$\bar{\bar{\ket{+}}}^i$ created from a single-qubit quantum state $\ket{+}^i$. Then, with the help of transversal quantum gates, local
measurements, and classical communication, quantum nodes apply non-transversal $\bar{\bar{H}}^i$ gate to the logical quantum state
$\bar{\bar{\mathcal{P}}}^i$, or in other words, achieve the realization of the logical quantum state
$\bar{\bar{H}}^i\bar{\bar{\mathcal{P}}}^i\bar{\bar{H}}^i$, see Fig.~\ref{fig:gate_teleportation_h}. During the gate teleportation protocol
information on the positions of the cheating quantum nodes is updated, see Sec.~\ref{subsec:gate_teleportation_protocol} for details.

We note that our MPQC protocol is information-theoretically secure, i.e., we make no assumptions on the computational power of the
non-adaptive active adversary, see Sec.~\ref{subsec:adversary}, but has an exponentially small probability of error inherited from the VHSS
protocol, i.e., $\kappa 2^{-\Omega(r)}$, where $r$ is the security parameter and $\kappa = n + \#\ancillas + \#H$, with $\#\ancillas$
standing for the number of ancillary quantum states required for the implementation of the quantum circuit $\mathcal{U}$ and $\#H$ standing
for the number of the $H$ gates~\footnote{Namely, $\kappa$ is equal to the number of times the VHSS protocol is invoked during the execution
of the MPQC protocol.}. Also, the aforementioned adversary in our MPQC protocol is limited only by the number of quantum nodes
$t < \frac{n}{4}$ it can corrupt, see Sec.~\ref{subsec:adversary}, which is a limitation derived from the Knill-Laflamme bound or the
quantum Singleton bound~\cite{knill_pra_55_900_1997}, see Ref.~\cite{smith_arxiv:quant-ph/0111030} for details. To be more specific, the
number of corrupted quantum nodes is constrained by the distance $d$ of the triply-even CSS QECC $\mathcal{C}_\mathrm{TE}$ as
$t \leq \left \lfloor \frac{d - 1}{2} \right \rfloor$, see Sec.~\ref{subsec:css_qeccs}. This constraint allows honest quantum nodes to
correct all the arbitrary quantum errors introduced by the $t < \frac{n}{4}$ cheating quantum nodes. Consequently, our MPQC protocol
satisfies the security requirements, i.e., \textit{correctness}, \textit{soundness}, and \textit{privacy}, which indeed hold with the
probability exponentially close to $1$ in the security parameter $r$, see Sec.~\ref{sec:security_mpqc_protocol}. Important to note that we
allow our MPQC protocol to abort at the end of computation if honest quantum nodes detect too many cheating quantum nodes during the
execution of the protocol, in a similar manner to Refs.~\cite{lipinska_pra_102_022405_2020,lipinska_arxiv:2004.10486v2}.

During the execution of the MPQC protocol, in addition to the $n^2$ single-qubit quantum states required for holding a share
$\bar{\bar{\mathcal{P}}}_i$, each quantum node $i$ uses $2n$ single-qubit ancillary quantum states to verify the encodings of the inputs
$\rho^1, \ldots, \rho^n$ by the VHSS protocol, see Sec.~\ref{subsec:vhss_protocol}, and $3n$ single-qubit ancillary quantum states to apply
a non-transversal $H$ gate with the gate teleportation technique involving verification of the ancillary logical quantum state
$\bar{\bar{\ket{+}}}^i$, see Sec.~\ref{subsec:gate_teleportation_protocol}, or to verify the ancillary logical quantum states
$\bar{\bar{\ket{0}}}^i$ which may be required for the implementation of the logical quantum circuit $\bar{\bar{\mathcal{U}}}$. Thus, in
total each quantum node requires $n^2 + 3n$ qubits for the implementation of the MPQC protocol. Finally, since the communication complexity
of the VHSS protocol per quantum node is $\mathcal{O}(nr^2)$ qubits, see Sec.~\ref{subsec:vhss_protocol}, the communication complexity of
our MPQC protocol per quantum node will be $\mathcal{O}\big((n + \#\ancillas + \#H)nr^2\big)$ qubits, see Sec.~\ref{sec:mpqc_protocol},
which is proportional to the total number of the VHSS protocol executions during the MPQC protocol.

% /////////////////////////////////////////////////////////////////////////////////////////////////////////////////////////////////////////
\section{
  \label{sec:assumptions_and_definitions}
  Assumptions and Definitions
}
% /////////////////////////////////////////////////////////////////////////////////////////////////////////////////////////////////////////

In this section, we overview our assumptions and definitions necessary for the construction of the MPQC protocol. In
Sec.~\ref{subsec:communication_channels}, we describe our assumptions on the classical and quantum communication channels as well as on the
broadcast channel, and in Sec.~\ref{subsec:adversary}, we describe our assumptions on the adversary. Next, in Sec.~\ref{subsec:css_qeccs},
we define properties common to any type of CSS QECCs, and finally, in Sec.~\ref{subsec:triply-even_css_qeccs}, we discuss a sub-class of
CSS QECCs, i.e., triply-even CSS QECCs.

% =========================================================================================================================================
\subsection{
  \label{subsec:communication_channels}
  Communication channels
}
% =========================================================================================================================================

In our MPQC protocol, we assume that all the quantum nodes have an access to the classical authenticated broadcast
channel~\cite{canetti_ieee_99_2_708-716_1999} (which is feasible if and only if $t < \frac{n}{3}$~\cite{lamport_acm_tpms_4_382-401_1982,
pease_jacm_27_228-234_1980}) and to the public source of randomness, the latter of which can be created with the help of the secure
multi-party classical computation~\cite{ben-or_stoc_88_1-10_1988,chaum_stoc_88_11-19_1988} (which is also feasible if and only if
$t < \frac{n}{3}$)~\footnote{We note that aforementioned constraints on the feasibility of the classical authenticated broadcast channel
and the public source of randomness do not cause any additional problems since we assume that only
$t < \frac{n}{4} \left(< \frac{n}{3}\right)$ quantum nodes are corrupted by the adversary, see Sec.~\ref{subsec:adversary}.}. Also, each
pair of quantum nodes is connected via the authenticated and private classical~\cite{canetti_ieee_17_219-233_2004} and
quantum~\cite{barum_ieee_43_449-458_2002} channels. Finally, we assume that each quantum node can perfectly process and store classical
and quantum information.

% =========================================================================================================================================
\subsection{
  \label{subsec:adversary}
  Adversary
}
% =========================================================================================================================================

In our MPQC protocol, we make no assumptions about the computational power of the adversary. Our non-adaptive~\footnote{Non-adaptive is the
adversary that chooses quantum nodes to corrupt before the MPQC protocol begins and remains with that choice.}, but active~\footnote{Active
is the adversary that is able to perform arbitrary quantum operations on the shares in the possession of the corrupted quantum nodes.}
adversary is limited only by the number of quantum nodes $t < \frac{n}{4}$ it can corrupt. The quantum nodes which are corrupted by the
adversary and therefore disobey the instructions of the MPQC protocol are called cheating quantum nodes. On the contrary, the quantum nodes
which are not corrupted by the adversary and obey the instructions of the MPQC protocol are called honest quantum nodes.

% =========================================================================================================================================
\subsection{
  \label{subsec:css_qeccs}
  CSS QECCs
}
% =========================================================================================================================================

Since in our MPQC protocol we consider a sub-class of general CSS QECCs~\cite{steane_prsla_452_2551-2577_1996,
calderbank_pra_54_1098_1996}, i.e., triply-even CSS QECCs $\mathcal{C}_\mathrm{TE}$~\cite{betsumiya_jlms_86_1_1-16_2012,
knill_arxiv:quant-ph/9610011}, we first define properties common to any type of CSS QECCs. Hereafter, $[n, k, d]$ stands for the distance
$d$ binary classical linear code that encodes $k$ bits into $n$ bits. General CSS QECC is defined through the two binary classical linear
codes denoted as $V$ and $W$, and these binary classical linear codes satisfy the following three conditions:
\begin{itemize}
  \item $V$ is an $[n, k_V, d_V]$ binary classical linear code that can correct $t_V \leq \left \lfloor \frac{d_V - 1}{2} \right \rfloor$
  bit errors.
  \item $W$ is an $[n, k_W, d_W]$ binary classical linear code that can correct $t_W \leq \left \lfloor \frac{d_W - 1}{2} \right \rfloor$
  bit errors.
  \item $V^\perp$ and $W$ satisfy $V^\perp \subseteq W$, where $V^\perp$ means the dual of the binary classical linear code $V$. Here,
  $V^\perp$ is an $[n, k_{V^\perp}, d_{V^\perp}]$ classical linear code that satisfies $k_{V^\perp} = n - k_V$.
\end{itemize}

These two binary classical linear codes generate an $[[n, k, d]]$ CSS QECC encoding $k$-qubit quantum state into $n$-qubit logical quantum
state, and where the constraint $k = k_V + k_W - n$ is satisfied. Such a CSS QECC can correct $t_V$ bit flip ($X$) errors and $t_W$ phase
flip ($Z$) errors, which leads to a CSS QECC with distance $d \geq \min (d_V, d_W)$ tolerating
$t \leq \left \lfloor \frac{d - 1}{2} \right \rfloor$ arbitrary quantum errors and $p \leq d - 1$ erasure quantum errors.

Since the VHSS protocol we employ in this paper works for any type of CSS QECCs encoding single-qubit quantum states into $n$-qubit logical
quantum state, see Sec.~\ref{subsec:vhss_protocol}, $k$ will always be equal to $1$, and therefore the encodings of the standard basis
``zero'' state and the Fourier basis ``plus'' state can be written as $\bar{\ket{0}} = \frac{1}{\sqrt{W^\bot}} \sum_{w \in W^\bot} \ket{w}$
and $\bar{\ket{+}} = \frac{1}{\sqrt{V}} \sum_{v \in V} \ket{v}$ correspondingly. Here individual codewords of the binary classical linear
codes $V$ and $W$ are denoted as $v$ and $w$ respectively.

Important to note that a CSS QECC generated by the two binary classical linear codes $V$ and $W$ may be denoted as a set
$V \cap \mathcal{F} W$ (where $\mathcal{F}$ stands for the Fourier transform), which means that a CSS QECC is a set of $n$-qubit logical
quantum states, which yield a codeword $v$ in $V$ when measured in the standard basis (also called $Z$ basis in the literature) and a
codeword $w$ in $W$ when measured in the Fourier basis (also called $X$ basis in the literature)~\cite{nielsen_chuang}.

Also, we emphasize that any type of CSS QECC allows transversal implementation of the $\mathrm{C}\text{-}X$ gate, while not any type of CSS
QECC allows transversal implementation of the other well-known quantum gates such as $H$ gate, $P$ gate, or $T$ gate. For example, a
sub-class of general CSS QECCs constructed from the two binary classical linear codes satisfying $V = W$ and called self-dual CSS
QECCs~\cite{preskill}, allows transversal implementation of $H$ gate, $P$ gate, and $\mathrm{C}\text{-}X$ gate, while does not allow
transversal implementation of the $T$ gate. Besides, we should note that in case of CSS QECCs, logical measurement can be implemented
transversally by local measurements of all the single-qubit quantum states comprising the $n$-qubit logical quantum state and the classical
communication.

Finally, let us mention the important property of CSS QECCs. The set of stabilizer generators $S$ of any CSS QECC can be divided into the
set of stabilizer generators consisting of only $X$ and $I$ (in this case, each stabilizer generator is denoted as $S^X_g$ and the entire
set is denoted as $S^X$) or only $Z$ and $I$ (in this case, each stabilizer generator is denoted as $S^Z_g$ and the entire set is denoted
as $S^Z$) operators in the tensor product representation, which permits independent correction of the bit flip ($X$) and the phase flip
($Z$) errors. As it happens, the Steane-type quantum error correction method~\cite{steane_prl_78_2252_1997} on the basis of which the VHSS
protocol is built, takes advantage of this fact~\cite{lipinska_pra_101_032332_2020}, see Sec.~\ref{subsec:vhss_protocol} for details.
Furthermore, the encoding of the standard basis ``zero'' state and the Fourier basis ``plus'' state in terms of the stabilizer generators
can be written as $\bar{\ket{0}} = \frac{1}{\sqrt{2^|S^X|}} \prod_{g \in S^X} (I + S^X_g) \ket{0}^{\otimes n}$ and
$\bar{\ket{+}} = \frac{1}{\sqrt{2^|S^Z|}} \prod_{g \in S^Z} (I + S^Z_g) \ket{+}^{\otimes n}$ correspondingly.

% =========================================================================================================================================
\subsection{
  \label{subsec:triply-even_css_qeccs}
  Triply-even CSS QECCs
}
% =========================================================================================================================================

To be comprehensive, we briefly describe a method of constructing the triply-even CSS QECCs
$\mathcal{C}_\mathrm{TE}$~\cite{betsumiya_jlms_86_1_1-16_2012,knill_arxiv:quant-ph/9610011}, which constitute a sub-class of general CSS
QECCs~\cite{steane_prsla_452_2551-2577_1996,calderbank_pra_54_1098_1996} and allow transversal implementation of the $T$ gate without any
Clifford corrections~\cite{rengaswamy_ieee_jsait_1_2_499-514_2020}. This is in contrast to the so called triorthogonal CSS QECCs, which
constitute an another sub-class of general CSS QECCs as well as comprise a super-class for the triply-even CSS QECCs
$\mathcal{C}_\mathrm{TE}$, and for which the transversal implementation of the $T$ gate requires additional Clifford
corrections~\cite{bravyi_pra_86_052329_2012}. We begin with the definition of the triply-even binary
matrices~\cite{betsumiya_jlms_86_1_1-16_2012} from which the triply-even CSS QECCs $\mathcal{C}_\mathrm{TE}$ can be
constructed~\cite{paetznick_arxiv:1410.5124}. Suppose the existence of two binary vectors $f$, $g \in \{0, 1\}^n$ with the Hamming weights
$|f|$ and $|g|$ respectively, and for which the entry-wise product $f \cdot g \in \{0, 1\}^n$ is defined. In this case, we call an
$m \times n$ binary matrix $G$ triorthogonal if for its rows $f_1, \ldots f_m \in \{0, 1\}^n$ following two conditions are satisfied:
% +++++++++++++++++++++++++++++++++++++++++++++++++++++++++++++++++++++++++++++++++++++++++++++++++++++++++++++++++++++++++++++++++++++++++
\begin{align}
|f_i \cdot f_j \cdot f_k| = 0 \pmod 2
\label{eq:condition_1}
\end{align}
% +++++++++++++++++++++++++++++++++++++++++++++++++++++++++++++++++++++++++++++++++++++++++++++++++++++++++++++++++++++++++++++++++++++++++
for all triples of rows $1 \leq i < j < k \leq m$,
% +++++++++++++++++++++++++++++++++++++++++++++++++++++++++++++++++++++++++++++++++++++++++++++++++++++++++++++++++++++++++++++++++++++++++
\begin{align}
|f_i \cdot f_j| = 0 \pmod 2
\label{eq:condition_2}
\end{align}
% +++++++++++++++++++++++++++++++++++++++++++++++++++++++++++++++++++++++++++++++++++++++++++++++++++++++++++++++++++++++++++++++++++++++++
for all pairs of rows $1 \leq i < j \leq m$. If in addition to the above two conditions the more restrictive constraint
% +++++++++++++++++++++++++++++++++++++++++++++++++++++++++++++++++++++++++++++++++++++++++++++++++++++++++++++++++++++++++++++++++++++++++
\begin{align}
|f_i \cdot f_j| = 0 \pmod 4
\end{align}
% +++++++++++++++++++++++++++++++++++++++++++++++++++++++++++++++++++++++++++++++++++++++++++++++++++++++++++++++++++++++++++++++++++++++++
is satisfied for all pairs of even weight rows $1 \leq i < j \leq l$, we call the binary matrix $G$ triply-even. The latter constraint
implies that $|f_i| = 0 \pmod 8$ is satisfied for all the even weight rows of the binary matrix $G$~\cite{paetznick_arxiv:1410.5124}.
Important to note that we assume the $m \times n$ binary matrix $G$ consisting of two submatrices: the one comprised of $l$ even weight
rows and denoted as $G_e$ (an $l \times n$ matrix) and the one comprised of $m - l$ odd weight rows and denoted as $G_o$ (an
$l - m \times n$ matrix).

With the above triply-even binary matrix $G$ at hand, one can construct corresponding triply-even CSS QECC $\mathcal{C}_\mathrm{TE}$ as
follows~\cite{bravyi_pra_86_052329_2012,paetznick_arxiv:1410.5124}. For each row of the binary matrix $G_e$, one defines an $X$ stabilizer
generator by mapping non-zero entries of the row to the $X$ operators (and zero entries of the row to the $I$ operators). Next, for each
row of the orthogonal complement of the triply-even binary matrix $G$, i.e., $G^\bot$, one defines a $Z$ stabilizer generator by mapping
non-zero entries of the row to the $Z$ operators (and zero entries of the row to the $I$ operators). Finally, each row of the binary matrix
$G_o$ corresponds to both the $\bar{X}$ and $\bar{Z}$ operators, if non-zero entries of the rows are mapped to the $X$ and $Z$ operators
respectively (and zero entries of the rows to the $I$ operators in both cases).

Let us also mention about the minimum distance of the triply-even CSS QECCs $\mathcal{C}_\mathrm{TE}$ constructed above. If we denote the
linear span of all the rows of the binary matrices $G_e$ ($G_e^\bot$) and $G^\bot$ as $\mathcal{G}_e$ ($\mathcal{G}_e^\bot$) and
$\mathcal{G}^\bot$ respectively, in case of the triorthogonal CSS QECCs, and consequently the triply-even CSS QECCs
$\mathcal{C}_\mathrm{TE}$, the condition $\mathcal{G}_e \subseteq \mathcal{G}^\bot$ is satisfied and this fact will automatically imply
the relation $d_Z \leq d_X$, where $d_Z$ and $d_X$ mean the distances against the phase flip $Z$ and bit flip $X$ errors
respectively~\cite{bravyi_pra_86_052329_2012,paetznick_arxiv:1410.5124}. Therefore, the minimum distance of the triply-even CSS QECC
$\mathcal{C}_\mathrm{TE}$ can be defined as the minimum weight of any non-trivial $\bar{Z}$ operator:
$\displaystyle d = \min_{f \in \mathcal{G}_e^\bot \setminus \mathcal{G}^\bot} |f|$~\cite{bravyi_pra_86_052329_2012,
nezami_pra_106_012437_2022}.

Eventually, to introduce an essential property of the triply-even CSS QECCs $\mathcal{C}_\mathrm{TE}$, i.e., a transversal implementation
of the $T$ gate without any Clifford corrections, we define the weight $|S|$ of a stabilizer generator $S$ as the number of terms not-equal
to $I$ in the tensor product representation. Actually, according to the Ref.~\cite{rengaswamy_ieee_jsait_1_2_499-514_2020}, a CSS QECC
allows transversal implementation of the $T$ gate without any Clifford corrections if and only if the binary matrix $G$ is triorthogonal,
i.e., Eqs.~\ref{eq:condition_1} and \ref{eq:condition_2} are satisfied, and the weight of all the stabilizer generators $S^X$ is multiple
of eight: $|S^X| = 0 \pmod 8$~\footnote{This statement is indeed consistent with the claim that the condition $|f_i| = 0 \pmod 8$ is
satisfied for all the even weight rows $1 \leq i \leq l$ of the triply-even binary matrix $G$.}. An example of such a CSS QECC is
$[[15, 1, 3]]$ CSS QECC~\cite{knill_arxiv:quant-ph/9610011}, as well as $[[49, 1, 5]]$ CSS QECC~\cite{bravyi_pra_86_052329_2012}.

As we can see from the above discussions, triply-even CSS QECCs $\mathcal{C}_\mathrm{TE}$ allow transversal implementation of $X$ gate, $Z$
gate, $\mathrm{C}\text{-}X$ gate (since any type of CSS QECC allows transversal implementation of these quantum gates), and $T$ gate.
Therefore, only the $H$ gate in the chosen universal set of quantum gates ($X$ gate, $Z$ gate, $\mathrm{C}\text{-}X$ gate, $T$ gate, and
$H$ gate) is not transversal and needs to be implemented by the gate teleportation technique, see
Sec.~\ref{subsec:gate_teleportation_protocol}, since in case of the triply-even CSS QECCs $\mathcal{C}_\mathrm{TE}$ binary classical linear
codes $V$ and $W$ do not satisfy $V = W$.

% /////////////////////////////////////////////////////////////////////////////////////////////////////////////////////////////////////////
\section{
  \label{sec:mpqc_subroutines}
  Subroutines of the MPQC protocol
}
% /////////////////////////////////////////////////////////////////////////////////////////////////////////////////////////////////////////

In this section, we describe subroutines used as building blocks in the construction of the MPQC protocol. In
Sec.~\ref{subsec:vhss_protocol}, we review the VHSS protocol used during the \textit{sharing}, \textit{verification}, and
\textit{reconstruction} phases of the MPQC protocol, and in Sec.~\ref{subsec:gate_teleportation_protocol}, we outline the gate
teleportation technique necessary for the implementation of the $H$ gate, which is non-transversal in case of triply-even CSS QECCs used in
our construction of the MPQC protocol.

% =========================================================================================================================================
\subsection{
  \label{subsec:vhss_protocol}
  Outline of the VHSS protocol
}
% =========================================================================================================================================

% -----------------------------------------------------------------------------------------------------------------------------------------
\begin{figure*}[htb]
  \begin{center}
% !!!!!!!!!!!!!!!!!!!!!!!!!!!!!!!!!!!!!!!!!!!!!!!!!!!!!!!!!!!!!!!!!!!!!!!!!!!!!!!!!!!!!!!!!!!!!!!!!!!!!!!!!!!!!!!!!!!!!!!!!!!!!!!!!!!!!!!!!
    \includegraphics[width=2.0\columnwidth]{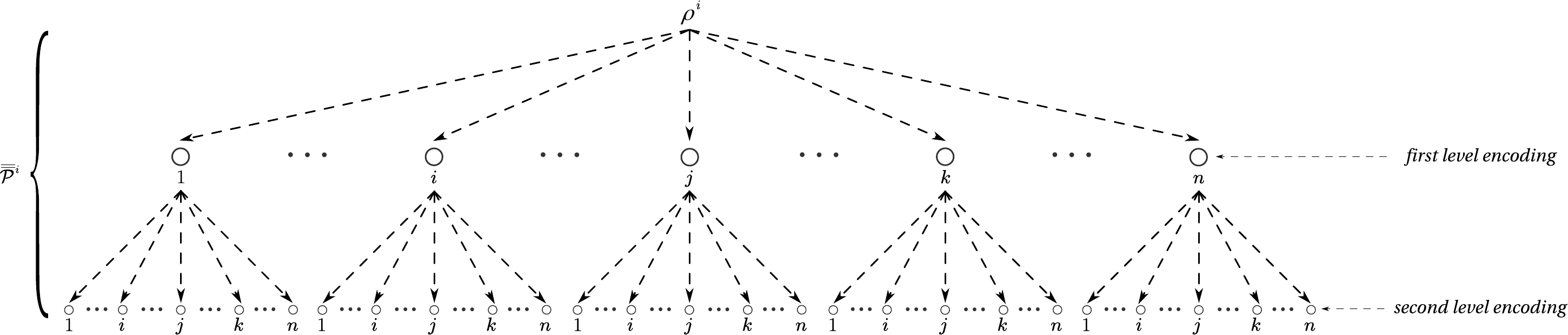}
% !!!!!!!!!!!!!!!!!!!!!!!!!!!!!!!!!!!!!!!!!!!!!!!!!!!!!!!!!!!!!!!!!!!!!!!!!!!!!!!!!!!!!!!!!!!!!!!!!!!!!!!!!!!!!!!!!!!!!!!!!!!!!!!!!!!!!!!!!
    \caption{
      Schematic picture of the sharing phase of the VHSS protocol during which a single-qubit input quantum state $\rho^i$ from the dealer
      $D^i$ undergoes the \textit{first level encoding} and the \textit{second level encoding} and eventually the global logical quantum
      state $\bar{\bar{\mathcal{P}}}^i$ is created. Each circle represents a single-qubit quantum state.
    }
% !!!!!!!!!!!!!!!!!!!!!!!!!!!!!!!!!!!!!!!!!!!!!!!!!!!!!!!!!!!!!!!!!!!!!!!!!!!!!!!!!!!!!!!!!!!!!!!!!!!!!!!!!!!!!!!!!!!!!!!!!!!!!!!!!!!!!!!!!
    \label{fig:vhss_sharing}
% !!!!!!!!!!!!!!!!!!!!!!!!!!!!!!!!!!!!!!!!!!!!!!!!!!!!!!!!!!!!!!!!!!!!!!!!!!!!!!!!!!!!!!!!!!!!!!!!!!!!!!!!!!!!!!!!!!!!!!!!!!!!!!!!!!!!!!!!!
  \end{center}
\end{figure*}
% -----------------------------------------------------------------------------------------------------------------------------------------

% -----------------------------------------------------------------------------------------------------------------------------------------
\begin{figure}[b]
  \begin{center}
% !!!!!!!!!!!!!!!!!!!!!!!!!!!!!!!!!!!!!!!!!!!!!!!!!!!!!!!!!!!!!!!!!!!!!!!!!!!!!!!!!!!!!!!!!!!!!!!!!!!!!!!!!!!!!!!!!!!!!!!!!!!!!!!!!!!!!!!!!
    \includegraphics[width=0.8\columnwidth]{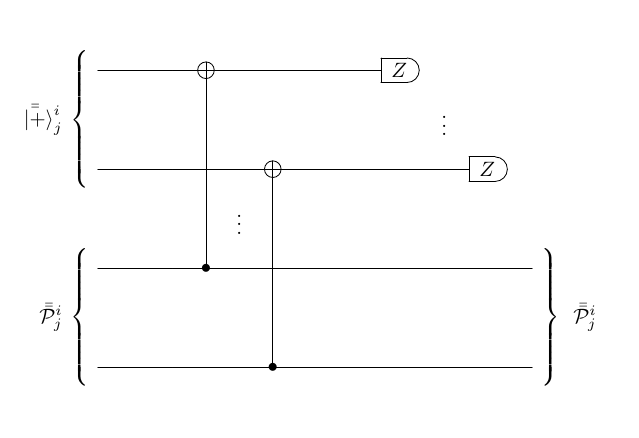}
% !!!!!!!!!!!!!!!!!!!!!!!!!!!!!!!!!!!!!!!!!!!!!!!!!!!!!!!!!!!!!!!!!!!!!!!!!!!!!!!!!!!!!!!!!!!!!!!!!!!!!!!!!!!!!!!!!!!!!!!!!!!!!!!!!!!!!!!!!
    \caption{
      Fragment of the logical quantum circuit $\bar{\bar{\mathcal{U}}}$ in which quantum node $j$ propagates arbitrary quantum errors in
      the share $\bar{\bar{\mathcal{P}}}^i_j$ (if any) to the ancillary share $\bar{\bar{\ket{+}}}^i_j$ to detect the bit flip $X$ errors.
      Note that the quantum gate and the logical measurement presented in the fragment of the logical quantum circuit
      $\bar{\bar{\mathcal{U}}}$ can be implemented transversally in case of any CSS QECC.
    }
% !!!!!!!!!!!!!!!!!!!!!!!!!!!!!!!!!!!!!!!!!!!!!!!!!!!!!!!!!!!!!!!!!!!!!!!!!!!!!!!!!!!!!!!!!!!!!!!!!!!!!!!!!!!!!!!!!!!!!!!!!!!!!!!!!!!!!!!!!
    \label{fig:bit_flip_errors_detection}
% !!!!!!!!!!!!!!!!!!!!!!!!!!!!!!!!!!!!!!!!!!!!!!!!!!!!!!!!!!!!!!!!!!!!!!!!!!!!!!!!!!!!!!!!!!!!!!!!!!!!!!!!!!!!!!!!!!!!!!!!!!!!!!!!!!!!!!!!!
  \end{center}
\end{figure}
% -----------------------------------------------------------------------------------------------------------------------------------------

% -----------------------------------------------------------------------------------------------------------------------------------------
\begin{table*}[htb]
% !!!!!!!!!!!!!!!!!!!!!!!!!!!!!!!!!!!!!!!!!!!!!!!!!!!!!!!!!!!!!!!!!!!!!!!!!!!!!!!!!!!!!!!!!!!!!!!!!!!!!!!!!!!!!!!!!!!!!!!!!!!!!!!!!!!!!!!!!
  \caption{
    Outline of the VHSS protocol.
  }
% !!!!!!!!!!!!!!!!!!!!!!!!!!!!!!!!!!!!!!!!!!!!!!!!!!!!!!!!!!!!!!!!!!!!!!!!!!!!!!!!!!!!!!!!!!!!!!!!!!!!!!!!!!!!!!!!!!!!!!!!!!!!!!!!!!!!!!!!!
  \begin{ruledtabular}
    \begin{tabularx}{\linewidth}{X}
% !!!!!!!!!!!!!!!!!!!!!!!!!!!!!!!!!!!!!!!!!!!!!!!!!!!!!!!!!!!!!!!!!!!!!!!!!!!!!!!!!!!!!!!!!!!!!!!!!!!!!!!!!!!!!!!!!!!!!!!!!!!!!!!!!!!!!!!!!
      \\
      \textbf{Input:} Private single-qubit quantum state $\rho^i$ (or $\ket{0}^i$, $\ket{+}^i$) from the dealer $D^i$ and an agreement on a
      particular $\mathcal{C}_\mathrm{TE}$. \\ \\
      \textbf{Output:} At the end of the verification phase, each quantum node $j = 1, \cdots, n$ holds a share
      $\bar{\bar{\mathcal{P}}}^i_j$ (or $_v\!\bar{\bar{\ket{0}}}^i_j$, $_v\!\bar{\bar{\ket{+}}}^i_j$) of the jointly verified logical
      quantum state $\bar{\bar{\mathcal{P}}}^i$ (or $_v\!\bar{\bar{\ket{0}}}^i$, $_v\!\bar{\bar{\ket{+}}}^i$) (and if required, quantum
      nodes are also able to confirm that what they hold is definitely a logical quantum state $_v\!\bar{\bar{\ket{0}}}^i$,
      $_v\!\bar{\bar{\ket{+}}}^i$) and a public set $B$.
      \\
% !!!!!!!!!!!!!!!!!!!!!!!!!!!!!!!!!!!!!!!!!!!!!!!!!!!!!!!!!!!!!!!!!!!!!!!!!!!!!!!!!!!!!!!!!!!!!!!!!!!!!!!!!!!!!!!!!!!!!!!!!!!!!!!!!!!!!!!!!
      \begin{enumerate}
        \item \textbf{Sharing:} Quantum nodes jointly create logical quantum state $\bar{\bar{\mathcal{P}}}^i$ (or
        $_v\!\bar{\bar{\ket{0}}}^i$, $_v\!\bar{\bar{\ket{+}}}^i$) by encoding and sharing input $\rho^i$ (or $\ket{0}^i$, $\ket{+}^i$)
        among all the $n$ quantum nodes. At the end of the sharing phase, each quantum node holds $n$ single-qubit quantum states coming
        from every other quantum node.
        \begin{enumerate}
          \item Dealer $D^i$ encodes his input $\rho^i$ (or $\ket{0}^i$, $\ket{+}^i$) into the $n$-qubit logical quantum state by using
          $\mathcal{C}_\mathrm{TE}$ and shares it among all the quantum nodes (including himself). We call this procedure the
          \textit{first level encoding}.
          \item Then, each quantum node $j = 1, \cdots, n$ one more time encodes a single-qubit quantum state obtained from the dealer
          $D^i$ into the $n$-qubit logical quantum state by using $\mathcal{C}_\mathrm{TE}$ and shares it among all the quantum nodes
          (including himself). We call this procedure the \textit{second level encoding}.
        \end{enumerate}
        \item \textbf{Verification:} Quantum nodes jointly verify that the input $\rho^i$ (or $\ket{0}^i$, $\ket{+}^i$) from the dealer
        $D^i$ is properly encoded and shared, and the valid logical quantum state $\bar{\bar{\mathcal{P}}}^i$ (or
        $_v\!\bar{\bar{\ket{0}}}^i$, $_v\!\bar{\bar{\ket{+}}}^i$) is created. Let us call this procedure as \textbf{verification of
        $\rho^i$}. Also, if required, quantum nodes jointly confirm that the input from the dealer $D^i$ is exactly $\ket{0}^i$
        ($\ket{+}^i$). Let us call this procedure as \textbf{confirmation of $\ket{0}^i$ ($\ket{+}^i$)}.
        \begin{enumerate}
          \item \textbf{Verification of $\rho^i$:} Quantum nodes create ancillary logical quantum states $\bar{\bar{\ket{0}}}^i$ and
          $\bar{\bar{\ket{+}}}^i$ with the same method as they created logical quantum state $\bar{\bar{\mathcal{P}}}^i$ in the sharing
          phase, propagate arbitrary quantum errors (if any) in the logical quantum state $\bar{\bar{\mathcal{P}}}^i$ to these ancillary
          logical quantum states, logically measure them in the appropriate basis, and decode the results of these logical measurements to
          find arbitrary quantum errors in the logical quantum state $\bar{\bar{\mathcal{P}}}^i$ (if any).
          \item \textbf{Confirmation of $\ket{0}^i$ ($\ket{+}^i$):} Quantum nodes create ancillary logical quantum states
          $\bar{\bar{\ket{0}}}^i$ ($\bar{\bar{\ket{+}}}^i$) with the same method as they created logical quantum state
          $\bar{\bar{\mathcal{P}}}^i$ in the sharing phase, propagate arbitrary quantum errors (if any) in the logical quantum state
          $_v\!\bar{\bar{\ket{0}}}^i$ ($_v\!\bar{\bar{\ket{+}}}^i$) to these ancillary logical quantum states, and logically measure them
          in the standard (Fourier) basis. Finally, quantum nodes decode the results of the logical measurements and publicly check whether
          they correspond to the $\ket{0}^i$ ($\ket{+}^i$).
          \item During the verification phase, quantum nodes jointly construct a public set $B$, which records all the arbitrary quantum
          errors introduced by the dealer $D^i$ and by the cheating quantum nodes.
          \item If at the end of the verification phase $|B| \leq t$ is satisfied, the dealer $D^i$ passes the verification phase, and the
          VHSS protocol continues to the reconstruction phase. On the other hand, if $|B| > t$ is satisfied the VHSS protocol aborts.
        \end{enumerate}
        \item \textbf{Reconstruction:} Reconstructor $R^j$ performs the following quantum operations on the single-qubit quantum states
        collected from the other quantum nodes and at the end of the reconstruction phase obtains output $\omega^j = \rho^i$.
        \begin{enumerate}
          \item Reconstructor $R^j$ identifies all the arbitrary quantum errors in each $n$-qubit logical quantum state originally encoded
          and shared by the quantum node $k \notin B$, by using $\mathcal{C}_\mathrm{TE}$. After that, the reconstructor $R^j$ decodes all
          the $n$-qubit logical quantum states with $t \leq \left \lfloor \frac{d - 1}{2} \right \rfloor$ arbitrary quantum errors. This is
          the \textit{second level decoding}. Otherwise, the reconstructor $R^j$ adds quantum node $k$ to the public set $B$.
          \item From the single-qubit quantum states obtained during the \textit{second level decoding}, the reconstructor $R^j$ randomly
          chooses $n - 2t$ single-qubit quantum states, each originally encoded and shared by the quantum node $k \notin B$, performs
          erasure recovery by using $\mathcal{C}_\mathrm{TE}$ and by decoding obtains output $\omega^j = \rho^i$. This is the \textit{first
          level decoding}.
        \end{enumerate}
      \end{enumerate}
% !!!!!!!!!!!!!!!!!!!!!!!!!!!!!!!!!!!!!!!!!!!!!!!!!!!!!!!!!!!!!!!!!!!!!!!!!!!!!!!!!!!!!!!!!!!!!!!!!!!!!!!!!!!!!!!!!!!!!!!!!!!!!!!!!!!!!!!!!
    \end{tabularx}
  \end{ruledtabular}
% !!!!!!!!!!!!!!!!!!!!!!!!!!!!!!!!!!!!!!!!!!!!!!!!!!!!!!!!!!!!!!!!!!!!!!!!!!!!!!!!!!!!!!!!!!!!!!!!!!!!!!!!!!!!!!!!!!!!!!!!!!!!!!!!!!!!!!!!!
  \label{table:vhss_protocol}
% !!!!!!!!!!!!!!!!!!!!!!!!!!!!!!!!!!!!!!!!!!!!!!!!!!!!!!!!!!!!!!!!!!!!!!!!!!!!!!!!!!!!!!!!!!!!!!!!!!!!!!!!!!!!!!!!!!!!!!!!!!!!!!!!!!!!!!!!!
\end{table*}
% -----------------------------------------------------------------------------------------------------------------------------------------

An important ingredient required for the construction of our MPQC protocol based on a technique of quantum error correction is the VHSS
protocol, which was recently introduced in Ref.~\cite{lipinska_pra_101_032332_2020}. First of all, quantum nodes participating in the MPQC
protocol use the VHSS protocol to encode and share a single-qubit input quantum state $\rho^i$ among all the $n$ quantum nodes in a
verifiable way. We note that the VHSS protocol in Ref.~\cite{lipinska_pra_101_032332_2020} is applicable to any type of CSS QECC and that
if the minimum distance of the underlying CSS QECC is $d$, the VHSS protocol tolerates
$t \leq \left \lfloor \frac{d - 1}{2} \right \rfloor$ cheating quantum nodes corrupted by the adversary described in
Sec.~\ref{subsec:adversary}. As it was already mentioned in Sec.~\ref{sec:mpqc_summary}, this constraint indeed allows honest quantum nodes
to correct all the arbitrary quantum errors introduced by the $t < \frac{n}{4}$ cheating quantum nodes. To be more specific, the VHSS
protocol is information-theoretically secure and satisfies the security requirements, i.e., \textit{soundness}, \textit{completeness}, and
\textit{secrecy}, which hold with the probability exponentially close to $1$ in the security parameter $r$. Namely, the verification
performed by using the VHSS protocol has the probability of error $2^{-\Omega(r)}$. The detailed security proof can be found in
Ref.~\cite{lipinska_pra_101_032332_2020}.

First, let us describe the VHSS protocol itself. In the sharing phase of the VHSS protocol, some quantum node $i$ acting as a dealer $D^i$
encodes his input $\rho^i$ into the $n$-qubit logical quantum state by using some CSS QECC (some triply-even CSS QECC
$\mathcal{C}_\mathrm{TE}$ in our case) on which all the quantum nodes have an agreement and shares it among all the quantum nodes
(including himself). We call this procedure the \textit{first level encoding}. Then, each quantum node $i$ one more time encodes a
single-qubit quantum state obtained from the dealer $D^i$ into the $n$-qubit logical quantum state by using the same CSS QECC (the same
triply-even CSS QECC $\mathcal{C}_\mathrm{TE}$ in our case) and one more time shares it among all the quantum nodes (including himself). We
call this procedure the \textit{second level encoding}. Eventually, quantum nodes jointly possess logical quantum state
$\bar{\bar{\mathcal{P}}}^i$ (or global logical quantum state $\bar{\bar{\mathcal{P}}}$ if all the $n$ quantum nodes participating in the
MPQC protocol have finished the sharing phase of the VHSS protocol), see Fig.~\ref{fig:vhss_sharing}.

In the verification phase of the VHSS protocol, quantum nodes jointly verify that the quantum state $\bar{\bar{\mathcal{P}}}^i$ in their
possession is for sure a valid logical quantum state encoded by the aforementioned triply-even CSS QECC $\mathcal{C}_\mathrm{TE}$. To be
more specific, quantum nodes publicly check that there are $t \leq \left \lfloor \frac{d - 1}{2} \right \rfloor$ arbitrary quantum errors
introduced by the dealer $D^i$ during the procedure of the \textit{first level encoding}, which will also mean that the dealer $D^i$ is
honest. For that purpose, first, quantum nodes jointly prepare ancillary logical quantum states $\bar{\bar{\ket{0}}}^i$ (to detect the
phase flip $Z$ errors) or $\bar{\bar{\ket{+}}}^i$ (to detect the bit flip $X$ errors), which are generated in the same way as the logical
quantum state $\bar{\bar{\mathcal{P}}}^i$ but from the single-qubit input quantum states $\ket{0}^i$ or $\ket{+}^i$ respectively. Then,
quantum nodes propagate arbitrary quantum errors in the logical quantum state $\bar{\bar{\mathcal{P}}}^i$ (if any) to these ancillary
logical quantum states by means of the transversal application of the $\overbar{\overbar{\mathrm{C}\text{-}X}}^i$ gate (superscript $i$
means that the logical quantum gate is applied between the logical quantum states initially created from the single-qubit input quantum
states $\rho^i$, $\ket{0}^i$, or $\ket{+}^i$) to their shares. Next, quantum nodes logically measure the ancillary logical quantum states
in the appropriate basis, and finally, by decoding the results of these logical measurements find arbitrary quantum errors in the logical
quantum state $\bar{\bar{\mathcal{P}}}^i$ (if any). The detailed procedure of the bit flip $X$ errors detection is shown in
Fig.~\ref{fig:bit_flip_errors_detection}.

Actually, the above procedure is an extension of the Steane-type quantum error correction method introduced in
Ref.~\cite{steane_prl_78_2252_1997} and stands for a single iteration in the verification phase of the VHSS protocol which contains
$r^2 + 2r$ of such iterations. To be more specific, there are $r$ iterations to check the bit flip $X$ errors (where each iteration spends
single ancillary logical quantum state $\bar{\bar{\ket{+}}}^i$), $r$ iterations to check the phase flip $Z$ errors (where each iteration
spends single ancillary logical quantum state $\bar{\bar{\ket{0}}}^i$), and $r$ additional iterations for each ancillary logical quantum
state $\bar{\bar{\ket{0}}}^i$ to check it for the bit flip $X$ errors (where each iteration indeed spends single ancillary logical quantum
state $\bar{\bar{\ket{+}}}^i$). Obviously, this procedure requires a workspace of $3n$ qubits per quantum node, i.e., a workspace of $n$
qubits per quantum node for each of the logical quantum states $\bar{\bar{\mathcal{P}}}^i$, $\bar{\bar{\ket{0}}}^i$, and
$\bar{\bar{\ket{+}}}^i$, which need to be stored simultaneously during the verification phase of the VHSS protocol, see
Ref.~\cite{lipinska_pra_101_032332_2020} for the details.

Throughout the verification phase of the VHSS protocol, quantum nodes jointly construct a public set of apparent cheaters $B$, which
records all the arbitrary quantum errors introduced by the dealer $D^i$ during the procedure of the \textit{first level encoding} and by
the cheating quantum nodes during the procedure of the \textit{second level encoding}. This allows identification of the cheating quantum
nodes with probability exponentially close to $1$ in the security parameter $r$, i.e., the probability of error is $2^{-\Omega(r)}$. Note
that it is impossible to distinguish arbitrary quantum errors introduced by the dealer $D^i$ from those introduced by the cheating quantum
nodes. Anyway, if at the end of the verification phase $|B| \leq t$ is satisfied, then the dealer $D^i$ passes the verification phase of
the VHSS protocol. In this case, a logical quantum state $\bar{\bar{\mathcal{P}}}^i$ in the possession of the quantum nodes can always be
reconstructed into the original input $\rho^i$ because arbitrary quantum errors introduced during the \textit{first level encoding} and
the \textit{second level encoding} can always be corrected by the triply-even CSS QECC $\mathcal{C}_\mathrm{TE}$, since we assume that
there are $t < \frac{n}{4}$ cheating quantum nodes and therefore $t \leq \left \lfloor \frac{d - 1}{2} \right \rfloor$ arbitrary quantum
errors at each level of encoding. On the other hand, if $|B| > t$ is satisfied the VHSS protocol aborts.

In the reconstruction phase of the VHSS protocol, some quantum node $j$ acting as a reconstructor $R^j$ collects all the single-qubit
quantum states from all the quantum nodes. Then, to correct arbitrary quantum errors introduced to the logical quantum state
$\bar{\bar{\mathcal{P}}}^i$ by the cheating quantum nodes after the verification phase and before the reconstruction phase, i.e., at the
\textit{second level encoding}, the reconstructor $R^j$ identifies arbitrary quantum errors in the $n$-qubit logical quantum states coming
from the quantum nodes not in the public set of apparent cheaters $B$ by using the triply-even CSS QECC $\mathcal{C}_\mathrm{TE}$ and
subsequently updates a public set of apparent cheaters $B$. Next, the reconstructor $R^j$ decodes all the $n$-qubit logical quantum states
in his possession that may contain $t \leq \left \lfloor \frac{d - 1}{2} \right \rfloor$ arbitrary quantum errors. We call this procedure
the \textit{second level decoding}. After that, from the single-qubit quantum states obtained during the \textit{second level decoding},
the reconstructor $R^j$ randomly chooses $n - 2t$ single-qubit quantum states originally encoded and shared by the quantum nodes which are
not in the public set of apparent cheaters $B$, performs erasure recovery by using the triply-even CSS QECC $\mathcal{C}_\mathrm{TE}$, and
finally obtains single-qubit output quantum state $\omega^j = \rho^i$. We note that the communication complexity of the VHSS protocol per
quantum node becomes $\mathcal{O}(nr^2)$ qubits, which is obvious considering that the quantum nodes send $n^2 - 1$ single qubit quantum
states $(r + 1)^2$ times in the process of the VHSS protocol execution~\cite{lipinska_pra_101_032332_2020}.

Actually, the VHSS protocol in Ref.~\cite{lipinska_pra_101_032332_2020} is also able to confirm that the single-qubit input quantum state
from the dealer $D^i$ is exactly $\ket{0}^i$ (or $\ket{+}^i$), i.e., that the quantum state $_v\!\bar{\bar{\ket{0}}}^i$
($_v\!\bar{\bar{\ket{+}}}^i$) (left subscript $v$ denotes the logical quantum state which the quantum nodes want to verify and confirm, and
is used to distinguish it from the ancillary logical quantum states $\bar{\bar{\ket{0}}}^i$ or $\bar{\bar{\ket{+}}}^i$ which are spent in
the verification phase of the VHSS protocol) in the possession of the quantum nodes is for sure a valid logical quantum state created from
the input $\ket{0}^i$ (or $\ket{+}^i$) and is definitely encoded by the triply-even CSS QECC $\mathcal{C}_\mathrm{TE}$, see
Refs.~\cite{crepeau_stoc_02_643-652_2002,lipinska_pra_101_032332_2020,smith_arxiv:quant-ph/0111030} for the details. To achieve that, all
the $r^2 + 2r$ iterations in the verification phase of the VHSS protocol are performed with the ancillary logical quantum states
$\bar{\bar{\ket{0}}}^i$ (or $\bar{\bar{\ket{+}}}^i$), which are indeed generated from the single-qubit input quantum states $\ket{0}^i$
(or $\ket{+}^i$)~\footnote{This is different from the VHSS protocol employed to check whether the quantum node $i$ is honest or not by
simply verifying the encoding of the input $\rho^i$.}. Also, after the logical measurements of these ancillary logical quantum states in
the standard (or Fourier) basis, quantum nodes publicly check that the twice decoded outcomes of the logical measurements correspond to the
$\ket{0}^i$ (or $\ket{+}^i$).

% =========================================================================================================================================
\subsection{
  \label{subsec:gate_teleportation_protocol}
  Outline of the gate teleportation protocol
}
% =========================================================================================================================================

% -----------------------------------------------------------------------------------------------------------------------------------------
\begin{figure}[b]
  \begin{center}
% !!!!!!!!!!!!!!!!!!!!!!!!!!!!!!!!!!!!!!!!!!!!!!!!!!!!!!!!!!!!!!!!!!!!!!!!!!!!!!!!!!!!!!!!!!!!!!!!!!!!!!!!!!!!!!!!!!!!!!!!!!!!!!!!!!!!!!!!!
    \includegraphics[width=\columnwidth]{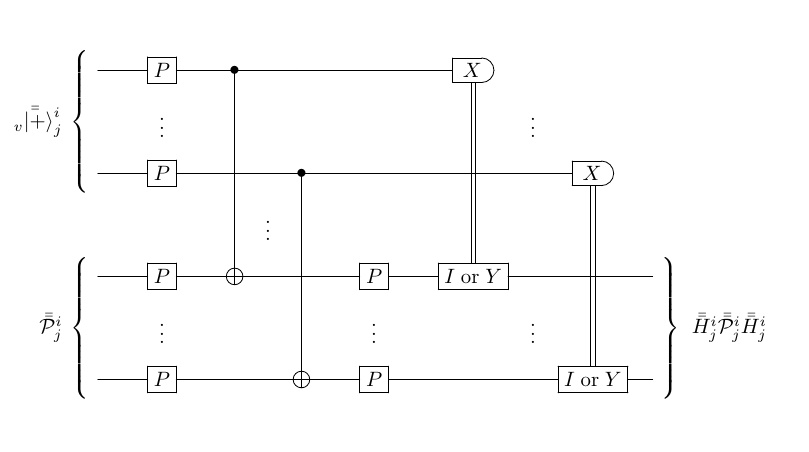}
% !!!!!!!!!!!!!!!!!!!!!!!!!!!!!!!!!!!!!!!!!!!!!!!!!!!!!!!!!!!!!!!!!!!!!!!!!!!!!!!!!!!!!!!!!!!!!!!!!!!!!!!!!!!!!!!!!!!!!!!!!!!!!!!!!!!!!!!!!
    \caption{
      Fragment of the logical quantum circuit $\bar{\bar{\mathcal{U}}}$ in which quantum node $j$ applies a non-transversal
      $\bar{\bar{H}}^i_j$ gate to the \textit{target} share $\bar{\bar{\mathcal{P}}}^i_j$ with the gate teleportation technique by taking
      advantage of the \textit{control} share $_v\!\bar{\bar{\ket{+}}}^i_j$. Note that the quantum gates and the logical measurement
      presented in the fragment of the logical quantum circuit $\bar{\bar{\mathcal{U}}}$ can be implemented transversally in case of the
      triply-even CSS QECCs $\mathcal{C}_\mathrm{TE}$.
    }
% !!!!!!!!!!!!!!!!!!!!!!!!!!!!!!!!!!!!!!!!!!!!!!!!!!!!!!!!!!!!!!!!!!!!!!!!!!!!!!!!!!!!!!!!!!!!!!!!!!!!!!!!!!!!!!!!!!!!!!!!!!!!!!!!!!!!!!!!!
    \label{fig:gate_teleportation_h}
% !!!!!!!!!!!!!!!!!!!!!!!!!!!!!!!!!!!!!!!!!!!!!!!!!!!!!!!!!!!!!!!!!!!!!!!!!!!!!!!!!!!!!!!!!!!!!!!!!!!!!!!!!!!!!!!!!!!!!!!!!!!!!!!!!!!!!!!!!
  \end{center}
\end{figure}
% -----------------------------------------------------------------------------------------------------------------------------------------

% -----------------------------------------------------------------------------------------------------------------------------------------
\begin{table*}[htb]
% !!!!!!!!!!!!!!!!!!!!!!!!!!!!!!!!!!!!!!!!!!!!!!!!!!!!!!!!!!!!!!!!!!!!!!!!!!!!!!!!!!!!!!!!!!!!!!!!!!!!!!!!!!!!!!!!!!!!!!!!!!!!!!!!!!!!!!!!!
  \caption{
    Outline of the gate teleportation protocol.
  }
% !!!!!!!!!!!!!!!!!!!!!!!!!!!!!!!!!!!!!!!!!!!!!!!!!!!!!!!!!!!!!!!!!!!!!!!!!!!!!!!!!!!!!!!!!!!!!!!!!!!!!!!!!!!!!!!!!!!!!!!!!!!!!!!!!!!!!!!!!
  \begin{ruledtabular}
    \begin{tabularx}{\linewidth}{X}
% !!!!!!!!!!!!!!!!!!!!!!!!!!!!!!!!!!!!!!!!!!!!!!!!!!!!!!!!!!!!!!!!!!!!!!!!!!!!!!!!!!!!!!!!!!!!!!!!!!!!!!!!!!!!!!!!!!!!!!!!!!!!!!!!!!!!!!!!!
      \\
      \textbf{Input:} Logical quantum state $\bar{\bar{\mathcal{P}}}^i$ verified by the VHSS protocol, ancillary logical quantum state
      $_v\!\bar{\bar{\ket{+}}}^i$ verified and confirmed by the VHSS protocol, and a public set $B$, see Sec.~\ref{subsec:vhss_protocol}.
      \\ \\
      \textbf{Output:} Non-transversal $\bar{\bar{H}}^i$ gate applied to the logical quantum state $\bar{\bar{\mathcal{P}}}^i$, i.e., a
      logical quantum state $\bar{\bar{H}}^i\bar{\bar{\mathcal{P}}}^i\bar{\bar{H}}^i$, and an updated public set $B$.
      \\
% !!!!!!!!!!!!!!!!!!!!!!!!!!!!!!!!!!!!!!!!!!!!!!!!!!!!!!!!!!!!!!!!!!!!!!!!!!!!!!!!!!!!!!!!!!!!!!!!!!!!!!!!!!!!!!!!!!!!!!!!!!!!!!!!!!!!!!!!!
      \begin{enumerate}
        \item \textbf{Quantum computation:} Each quantum node $j = 1, \cdots, n$ performs the following quantum operations on the $2n$
        single-qubit quantum states among which there are $n$ single-qubit quantum states comprising a share $\bar{\bar{\mathcal{P}}}^i_j$
        at the beginning of the gate teleportation protocol (called \textit{target} share hereafter), and $n$ single-qubit quantum states
        comprising a share $_v\!\bar{\bar{\ket{+}}}^i_j$ also at the beginning of the gate teleportation protocol (called \textit{control}
        share hereafter).
        \begin{enumerate}
          \item Quantum node $j$ applies transversal $\bar{\bar{P}}^i_j$ gate to both: \textit{target} and \textit{control} shares.
          \item Quantum node $j$ applies transversal $\overbar{\overbar{\mathrm{C}\text{-}X}}^i_j$ gate with \textit{control} share as the
          control and \textit{target} share as the target.
          \item Quantum node $j$ applies transversal $\bar{\bar{P}}^i_j$ gate to the \textit{target} share.
          \item Quantum node $j$ measures each single-qubit quantum state of the \textit{control} share in the Fourier basis and announces
          his measurement outcome using a classical authenticated broadcast channel, see Sec.~\ref{subsec:communication_channels}.
        \end{enumerate}
        \item \textbf{Classical computation:} The measurement outcomes announced by all the quantum nodes yield codewords in $W$ when
        rearranged into the groups in such a way that each group corresponds to the logical measurement outcome of the $n$-qubit logical
        quantum state (there are $n$ of them) originally encoded and shared by some quantum node $k$. Next, quantum nodes publicly check
        the positions of the arbitrary quantum errors by decoding the results of the logical measurements and consequently update the set
        $B$. Also, by decoding the codewords in $W$ twice, quantum nodes jointly identify whether the logical measurement results of the
        \textit{control} shares reconstruct to the single-qubit quantum states $\ket{+}^i$ or $\ket{-}^i$.
        \item \textbf{Correction:} According to the twice decoded outcomes of the logical measurements, each quantum node
        $j = 1, \cdots, n$ performs the following quantum operations on his \textit{target} share.
        \begin{itemize}
          \item If the twice decoded outcomes correspond to the $\ket{-}^i$, then the quantum node $j$ does nothing to his \textit{target}
          share.
          \item If the twice decoded outcomes correspond to the $\ket{+}^i$, then the quantum node $j$ transversally applies the
          $-i\bar{\bar{Y}}^i_j$ gate to his \textit{target} share.
        \end{itemize}
      \end{enumerate}
% !!!!!!!!!!!!!!!!!!!!!!!!!!!!!!!!!!!!!!!!!!!!!!!!!!!!!!!!!!!!!!!!!!!!!!!!!!!!!!!!!!!!!!!!!!!!!!!!!!!!!!!!!!!!!!!!!!!!!!!!!!!!!!!!!!!!!!!!!
    \end{tabularx}
  \end{ruledtabular}
% !!!!!!!!!!!!!!!!!!!!!!!!!!!!!!!!!!!!!!!!!!!!!!!!!!!!!!!!!!!!!!!!!!!!!!!!!!!!!!!!!!!!!!!!!!!!!!!!!!!!!!!!!!!!!!!!!!!!!!!!!!!!!!!!!!!!!!!!!
  \label{table:gate_teleportation_protocol}
% !!!!!!!!!!!!!!!!!!!!!!!!!!!!!!!!!!!!!!!!!!!!!!!!!!!!!!!!!!!!!!!!!!!!!!!!!!!!!!!!!!!!!!!!!!!!!!!!!!!!!!!!!!!!!!!!!!!!!!!!!!!!!!!!!!!!!!!!!
\end{table*}
% -----------------------------------------------------------------------------------------------------------------------------------------

Here we describe the gate teleportation technique which was first suggested in Ref.~\cite{gottesman_nature_402_390-393_1999}. The key idea
of the technique is to use a specially created ancillary quantum state as a \textit{control} quantum state, measure it with respect to the
appropriate basis, and apply necessary quantum correction to the \textit{target} quantum state depending on the measurement outcome. Gate
teleportation technique is frequently used for the fault-tolerant realization of the quantum gate that cannot be implemented transversally
once a particular QECC is chosen~\cite{gottesman_arxiv:quant-ph/0904.2557}. Since our MPQC protocol is constructed on the basis of the
triply-even CSS QECCs $\mathcal{C}_\mathrm{TE}$~\cite{betsumiya_jlms_86_1_1-16_2012,knill_arxiv:quant-ph/9610011}, the only quantum gate
which cannot be implemented transversally in the chosen universal set of quantum gates, i.e., $X$ gate, $Z$ gate, $T$ gate,
$\mathrm{C}\text{-}X$ gate, and $H$ gate, will be the $H$ gate~\cite{knill_arxiv:quant-ph/9610011}. Therefore, in our MPQC protocol, a
non-transversal $H$ gate needs to be implemented by the gate teleportation technique~\cite{knill_arxiv:quant-ph/9610011}.

The gate teleportation protocol implementing the non-transversal $H^i$ gate takes logical quantum state $\bar{\bar{\mathcal{P}}}^i$ and
ancillary logical quantum state $_v\!\bar{\bar{\ket{+}}}^i$ as an input, see Fig.~\ref{fig:gate_teleportation_h}. At this point, both of
these logical quantum states are already verified by using the VHSS protocol. In addition, quantum nodes has already jointly confirmed that
the ancillary logical quantum state $_v\!\bar{\bar{\ket{+}}}^i$ in their possession is definitely a logical version of the single-qubit
quantum state $\ket{+}^i$. Important to note that this can be achieved by using the VHSS protocol only, see
Sec.~\ref{subsec:vhss_protocol}. Here lies the main difference from the previous suggestion in Ref.~\cite{lipinska_pra_102_022405_2020} as
well as its reconsidered version in Ref.~\cite{lipinska_arxiv:2004.10486v2}, where the logical version of the ancillary ``magic'' state
$\frac{1}{\sqrt{2}}(\ket{0} + e^{i\pi/4}\ket{1})$, which is required for the implementation of the non-transversal $T$ gate with the gate
teleportation technique, cannot be verified by using the VHSS protocol only, and therefore an additional verification of the ancillary
logical ``magic'' state becomes vital~\cite{lipinska_pra_102_022405_2020,lipinska_arxiv:2004.10486v2}, see
Appendix~\ref{sec:magic_state_verification_protocol} for the details~\footnote{For the details of the protocol called ``verification of the
Clifford stabilizer states'' (VCSS) which was employed in the original version of the MPQC protocol for the verification of the ancillary
logical ``magic'' state see Appendix~\ref{sec:vcss_summary} and Appendix~\ref{sec:vcss_critics}}.

To perform the gate teleportation technique and apply a non-transversal $\bar{\bar{H}}^i$ gate to the logical quantum state
$\bar{\bar{\mathcal{P}}}^i$, in addition to the $n^2$ qubits required for holding a share $\bar{\bar{\mathcal{P}}}_i$, each quantum node
requires $3n$ qubits to verify and confirm the input ancillary logical quantum state $_v\!\bar{\bar{\ket{+}}}^i$ by using the VHSS
protocol, see Sec.~\ref{subsec:vhss_protocol}, and afterwards $n$ qubits to actually perform the gate teleportation technique. Therefore,
the communication complexity of the gate teleportation protocol is the same as of the VHSS protocol, i.e., $\mathcal{O}(nr^2)$ qubits per
quantum node.

The detailed procedure of the gate teleportation technique is shown in Fig.~\ref{fig:gate_teleportation_h}. To apply a non-transversal
$\bar{\bar{H}}^i$ gate to the logical quantum state $\bar{\bar{\mathcal{P}}}^i$, quantum nodes transversally apply
$\bar{\bar{P}}^i = \bar{\bar{T}}^i \circ \bar{\bar{T}}^i$ gate to their shares of both input logical quantum states
$\bar{\bar{\mathcal{P}}}^i$ and $_v\!\bar{\bar{\ket{+}}}^i$, then transversally apply $\overbar{\overbar{\mathrm{C}\text{-}X}}^i$ gate to
their shares, taking shares of the ancillary logical quantum state $_v\!\bar{\bar{\ket{+}}}^i$ as the \textit{control} shares and shares
of the logical quantum state $\bar{\bar{\mathcal{P}}}^i$ as the \textit{target} shares. Then, quantum nodes transversally apply
$\bar{\bar{P}}^i$ gate to the \textit{target} shares and logically measure the \textit{control} shares in the Fourier basis. Next, quantum
nodes decode the result of the logical measurement twice and publicly check whether this twice-decoded result corresponds to the
$\ket{+}^i$ or $\ket{-}^i$. Finally, if the result corresponds to the $\ket{-}^i$, then quantum nodes do nothing to their \textit{target}
shares, but if the result corresponds to the $\ket{+}^i$, then quantum nodes transversally apply the
$-i\bar{\bar{Y}}^i = -i\bar{\bar{P}}^i \circ \bar{\bar{X}}^i \circ \overbar{\overbar{P^\dag}}^i$ gate to their \textit{target} shares, see
Appendix~\ref{sec:h_gate_teleportation} for the detailed calculations. At the same time, quantum nodes update the public set of apparent
cheaters $B$, and if $|B| > t$ is satisfied, quantum nodes assume that the twice-decoded result of the logical measurement corresponds to
the $\ket{-}^i$, and do nothing to their \textit{target} shares. For the detailed procedure of the gate teleportation protocol see
Table~\ref{table:gate_teleportation_protocol}.

% /////////////////////////////////////////////////////////////////////////////////////////////////////////////////////////////////////////
\section{
  \label{sec:mpqc_protocol}
  Outline of the MPQC protocol
}
% /////////////////////////////////////////////////////////////////////////////////////////////////////////////////////////////////////////

% -----------------------------------------------------------------------------------------------------------------------------------------
\begin{table*}[htb]
% !!!!!!!!!!!!!!!!!!!!!!!!!!!!!!!!!!!!!!!!!!!!!!!!!!!!!!!!!!!!!!!!!!!!!!!!!!!!!!!!!!!!!!!!!!!!!!!!!!!!!!!!!!!!!!!!!!!!!!!!!!!!!!!!!!!!!!!!!
  \caption{
    Outline of the MPQC protocol.
  }
% !!!!!!!!!!!!!!!!!!!!!!!!!!!!!!!!!!!!!!!!!!!!!!!!!!!!!!!!!!!!!!!!!!!!!!!!!!!!!!!!!!!!!!!!!!!!!!!!!!!!!!!!!!!!!!!!!!!!!!!!!!!!!!!!!!!!!!!!!
  \begin{ruledtabular}
    \begin{tabularx}{\linewidth}{X}
% !!!!!!!!!!!!!!!!!!!!!!!!!!!!!!!!!!!!!!!!!!!!!!!!!!!!!!!!!!!!!!!!!!!!!!!!!!!!!!!!!!!!!!!!!!!!!!!!!!!!!!!!!!!!!!!!!!!!!!!!!!!!!!!!!!!!!!!!!
      \\
      \textbf{Input:} Private single-qubit quantum state $\rho^i$ from each quantum node $i$, agreement on a particular
      $\mathcal{C}_\mathrm{TE}$ and on a particular $\mathcal{U}$. \\ \\
      \textbf{Output:} In case of success, each quantum node $i$ possesses a private single-qubit quantum state $\omega^i$. In
      case of failure, the honest quantum nodes replace all the single-qubit quantum states in their possession with $\ket{0}$ and the MPQC
      protocol is aborted at the end of the computation.
      \\
% !!!!!!!!!!!!!!!!!!!!!!!!!!!!!!!!!!!!!!!!!!!!!!!!!!!!!!!!!!!!!!!!!!!!!!!!!!!!!!!!!!!!!!!!!!!!!!!!!!!!!!!!!!!!!!!!!!!!!!!!!!!!!!!!!!!!!!!!!
      \begin{enumerate}
        \item \textbf{Sharing:} For $i = 1, \cdots, n$, quantum nodes execute the sharing phase of the VHSS protocol with the quantum node
        $i$ acting as a dealer $D^i$ and the $\rho^i$ as an input, and jointly prepare logical quantum state $\bar{\bar{\mathcal{P}}}^i$,
        see Table~\ref{table:vhss_protocol}. When all the $n$ quantum nodes participating in the MPQC protocol have finished the sharing
        phase of the VHSS protocol, quantum nodes jointly possess a global logical quantum state $\bar{\bar{\mathcal{P}}}$.
        \item \textbf{Verification:} For $i = 1, \cdots, n$, quantum nodes execute the verification phase of the VHSS protocol with the
        quantum node $i$ acting as a dealer $D^i$ and jointly verify the logical quantum state $\bar{\bar{\mathcal{P}}}^i$, see
        Table~\ref{table:vhss_protocol}.
        \begin{enumerate}
          \item Quantum nodes jointly construct a public set $B^{i,j}$ which records all the arbitrary quantum errors introduced by the
          dealer $D^i$ and by the cheating quantum nodes during all the $n$ executions of the VHSS protocol. For $j = 1, \cdots, n$,
          if $|B^{i,j}| > t$ is satisfied, then quantum nodes add quantum node $j$ to the public set $B^i$.
          \item After all the $n$ executions of the VHSS protocol, quantum nodes jointly construct a global public set $B = \bigcup_i B^i$.
          If $|B| > t$ is satisfied, the \textit{abortion sequence} is invoked.
        \end{enumerate}
        \item \textbf{Computation:} Quantum nodes apply logical quantum gates ($\bar{\bar{X}}^i$ gate, $\bar{\bar{Z}}^i$ gate,
        $\bar{\bar{T}}^i$ gate, $\overbar{\overbar{\mathrm{C}\text{-}X}}^{i,j}$ gate, and $\bar{\bar{H}}^i$ gate) to the global logical
        quantum state $\bar{\bar{\mathcal{P}}}$ in a particular order specified by the logical quantum circuit $\bar{\bar{\mathcal{U}}}$.
        \begin{enumerate}
          \item For every transversal $\bar{\bar{X}}^i$ gate, $\bar{\bar{Z}}^i$ gate, or $\bar{\bar{T}}^i$ gate applied to the logical
          quantum state $\bar{\bar{\mathcal{P}}}^i$, each quantum node $j = 1, \cdots, n$ applies $X$ gates, $Z$ gates, or $T$ gates to
          the $n$ single-qubit quantum states comprising his share $\bar{\bar{\mathcal{P}}}^i_j$.
          \item For every transversal $\overbar{\overbar{\mathrm{C}\text{-}X}}^{i,j}$ gate applied between the logical quantum states
          $\bar{\bar{\mathcal{P}}}^i$ and $\bar{\bar{\mathcal{P}}}^j$, each quantum node $k = 1, \cdots, n$ applies
          $\mathrm{C}\text{-}X^{i,j}$ gates between the $n$ single-qubit quantum states comprising a share $\bar{\bar{\mathcal{P}}}^i_k$
          and the $n$ single-qubit quantum states comprising a share $\bar{\bar{\mathcal{P}}}^j_k$.
          \item For every non-transversal $\bar{\bar{H}}^i$ gate applied to the logical quantum state $\bar{\bar{\mathcal{P}}}^i$, quantum
          nodes take the following two actions:
          \begin{enumerate}
            \item Quantum nodes jointly create, then verify and confirm ancillary logical quantum state $_v\!\bar{\bar{\ket{+}}}^i$ by
            using the VHSS protocol, see Table~\ref{table:vhss_protocol}.
            \item Then, quantum nodes jointly perform the gate teleportation protocol with two input logical quantum states:
            $\bar{\bar{\mathcal{P}}}^i$ and $_v\!\bar{\bar{\ket{+}}}^i$, see Table~\ref{table:gate_teleportation_protocol}, and if at the
            end of the gate teleportation protocol execution $|B| > t$ is satisfied, the \textit{abortion sequence} is invoked.
          \end{enumerate}
          \item If the ancillary single-qubit quantum state $\ket{0}^i$ is required for the implementation of the quantum circuit
          $\mathcal{U}$, quantum nodes jointly create, then verify and confirm ancillary logical quantum state $_v\!\bar{\bar{\ket{0}}}^i$
          by using the VHSS protocol with the randomly chosen quantum node $i \notin B$ acting as a dealer $D^i$.
          \item If at any stage of the MPQC protocol execution $|B| > t$ is satisfied, the \textit{abortion sequence} is invoked.
        \end{enumerate}
        \item \textbf{Reconstruction:} Each quantum node $i = 1, \cdots, n$ executes the reconstruction phase of the VHSS protocol as a
        reconstructor $R^i$ after collecting all the single-qubit quantum states corresponding to his output logical quantum state
        $\bar{\bar{\Omega}}^i$ from the other quantum nodes.
        \begin{enumerate}
          \item Reconstructor $R^i$ identifies arbitrary quantum errors in each $n$-qubit logical quantum state originally encoded and
          shared by the quantum node $j \notin B$ during the second level encoding, by using $\mathcal{C}_\mathrm{TE}$. In parallel, the
          reconstructor $R^i$ creates another public set $\tilde{B}^{i,j}$ which records all the arbitrary quantum errors introduced by the
          cheating quantum nodes at the second level encoding and satisfies $\tilde{B}^{i,j} \subseteq B^{i,j}$. Then, the reconstructor
          $R^i$ decodes each $n$-qubit logical quantum state satisfying $|\tilde{B}^{i,j}| \leq t$. On the other hand, if some $n$-qubit
          logical quantum state originally encoded and shared by the quantum node $j \notin B$ during the second level encoding satisfies
          $|\tilde{B}^{i,j}| > t$, the reconstructor $R^i$ adds quantum node $j$ to the public set $B$.
          \item Reconstructor $R^i$ randomly chooses $n - 2t$ single-qubit quantum states, each originally encoded and shared by the
          quantum node $j \notin B$, performs erasure recovery by using $\mathcal{C}_\mathrm{TE}$, and by decoding obtains output
          $\omega^i$.
        \end{enumerate}
      \end{enumerate}
% !!!!!!!!!!!!!!!!!!!!!!!!!!!!!!!!!!!!!!!!!!!!!!!!!!!!!!!!!!!!!!!!!!!!!!!!!!!!!!!!!!!!!!!!!!!!!!!!!!!!!!!!!!!!!!!!!!!!!!!!!!!!!!!!!!!!!!!!!
    \end{tabularx}
  \end{ruledtabular}
% !!!!!!!!!!!!!!!!!!!!!!!!!!!!!!!!!!!!!!!!!!!!!!!!!!!!!!!!!!!!!!!!!!!!!!!!!!!!!!!!!!!!!!!!!!!!!!!!!!!!!!!!!!!!!!!!!!!!!!!!!!!!!!!!!!!!!!!!!
  \label{table:mpqc_protocol}
% !!!!!!!!!!!!!!!!!!!!!!!!!!!!!!!!!!!!!!!!!!!!!!!!!!!!!!!!!!!!!!!!!!!!!!!!!!!!!!!!!!!!!!!!!!!!!!!!!!!!!!!!!!!!!!!!!!!!!!!!!!!!!!!!!!!!!!!!!
\end{table*}
% -----------------------------------------------------------------------------------------------------------------------------------------

Here we describe our MPQC protocol in more detail. First of all, the entire MPQC protocol consists of \textit{sharing},
\textit{verification}, \textit{computation}, and \textit{reconstruction} phases, and has two sub-protocols as its building blocks, i.e.,
the VHSS protocol and the gate teleportation protocol. Let us see the entire flow of the MPQC protocol by closing up each phase and in
parallel explaining how the two sub-protocols are involved in the process.

\textbf{Sharing:} At this stage of the MPQC protocol, quantum nodes create global logical quantum state $\bar{\bar{\mathcal{P}}}$ by
executing the sharing phase of the VHSS protocol $n$ times. Each time quantum nodes execute the sharing phase of the VHSS protocol with the
quantum node $i$ acting as a dealer $D^i$ and the single-qubit quantum state $\rho^i$ as an input, they jointly prepare logical quantum
state $\bar{\bar{\mathcal{P}}}^i$, see Table~\ref{table:vhss_protocol}. Here, each quantum node $j$ requires a workspace of $n^2$ qubits
for holding his share of the global logical quantum state $\bar{\bar{\mathcal{P}}}$, i.e., a share $\bar{\bar{\mathcal{P}}}_j$. We also
note that this phase of the MPQC protocol has a communication complexity of $\mathcal{O}(n^2)$ qubits per quantum node, which is obvious
considering that at this stage of the MPQC protocol quantum nodes simply execute the sharing phase of the VHSS protocol $n$ times. For the
details see Table~\ref{table:mpqc_protocol}.

\textbf{Verification:} At this stage of the MPQC protocol, quantum nodes jointly verify that the quantum state $\bar{\bar{\mathcal{P}}}$ in
their possession is for sure a valid logical quantum state encoded by the triply-even CSS QECC $\mathcal{C}_\mathrm{TE}$ which is achieved
by executing the verification phase of the VHSS protocol $n$ times. Each time quantum nodes execute the verification phase of the VHSS
protocol with the quantum node $i$ acting as a dealer $D^i$, they jointly verify the logical quantum state $\bar{\bar{\mathcal{P}}}^i$ by
recording the positions of arbitrary quantum errors introduced the dealer $D^i$ at the first level encoding and by the cheating quantum
nodes at the second level encoding in a public set of apparent cheaters $B^i$, and in such a way check whether each quantum node $j$ is
honest or not, see Table~\ref{table:vhss_protocol}. After executing the verification phase of the VHSS protocol $n$ times, quantum nodes
jointly construct a global public set of apparent cheaters $B = \bigcup_i B^i$. If at the end of the verification phase $|B| \leq t$ is
satisfied, then quantum nodes proceed to the computation phase with their shares of the input global logical quantum state
$\bar{\bar{\mathcal{P}}}$, i.e., each quantum node $i$ holds a share $\bar{\bar{\mathcal{P}}}_i$. On the other hand, if $|B| > t$ is
satisfied, quantum nodes also proceed to the computation phase but the honest quantum nodes replace all the single-qubit quantum states in
their possession with $\ket{0}$ and the MPQC protocol is aborted at the end of the computation, see
Ref.~\cite{lipinska_pra_102_022405_2020} for the details. We call this procedure the \textit{abortion sequence}. This phase of the MPQC
protocol requires a workspace of $n^2 + 2n$ qubits per quantum node for the implementation, among which, $2n$ qubits are required for
holding ancillary logical quantum states $\bar{\bar{\ket{0}}}^i$ and $\bar{\bar{\ket{+}}}^i$ during the verification phase of the VHSS 
protocol. Also, we note that this phase of the MPQC protocol has a communication complexity of $\mathcal{O}(n^2r^2)$ qubits per quantum
node, since at this stage of the MPQC protocol quantum nodes simply execute the verification phase of the VHSS protocol $n$ times. For the
details see Table~\ref{table:mpqc_protocol}.

\textbf{Computation:} At this stage of the MPQC protocol, quantum nodes jointly perform logical quantum circuit $\bar{\bar{\mathcal{U}}}$
on the jointly verified global logical quantum state $\bar{\bar{\mathcal{P}}}$, and at the end of this stage quantum nodes will jointly
possess some output global logical quantum state $\bar{\bar{\Omega}}$, from which each quantum node $i$ can calculate his output logical
quantum state $\bar{\bar{\Omega}}^i = \mathrm{Tr}_{[n] \backslash i} (\bar{\bar{\Omega}})$. We note that the global public set of apparent
cheaters $B$ is cumulative throughout the entire MPQC protocol, namely, during the computation phase the global public set of apparent
cheaters $B$ is updated whenever the VHSS protocol or the gate teleportation protocol is invoked, see Table~\ref{table:mpqc_protocol}. If
at any stage of the MPQC protocol execution $|B| > t$ is satisfied, the honest quantum nodes replace all the single-qubit quantum states in
their possession with $\ket{0}$, and the MPQC protocol is aborted at the end of the computation. Otherwise, quantum nodes proceed to the
reconstruction phase with their shares of the output global logical quantum state $\bar{\bar{\Omega}}$. Application of the transversal
quantum gates, i.e., $\bar{\bar{X}}^i$ gate, $\bar{\bar{Z}}^i$ gate, $\bar{\bar{T}}^i$ gate, and
$\overbar{\overbar{\mathrm{C}\text{-}X}}^{i,j}$ gate (superscript $i,j$ means that the non-logical version of the quantum gate is applied
between the quantum wires $i$ and $j$ of the quantum circuit $\mathcal{U}$, where the quantum wire $i$ acts as a control and the quantum
wire $j$ acts as a target) does not require any additional workspace. On the other hand, whenever the implementation of the logical quantum
circuit $\bar{\bar{\mathcal{U}}}$ requires an ancillary logical quantum state or whenever the non-transversal $\bar{\bar{H}}^i$ gate is
applied, each quantum node will require an additional workspace of $3n$ qubits to verify and confirm the ancillary logical quantum states
$\bar{\bar{\ket{0}}}^i$ or $\bar{\bar{\ket{+}}}^i$ respectively by using the VHSS protocol, see
Table~\ref{table:gate_teleportation_protocol}. Therefore, this phase of the MPQC protocol requires a workspace of $n^2 + 3n$ qubits per
quantum node for the implementation and has a communication complexity of $\mathcal{O}\big((\#\ancillas + \#H)nr^2\big)$ qubits per quantum
node, which is easily evaluated from the number of times the VHSS protocol is invoked during the computation phase. For the details see
Table~\ref{table:mpqc_protocol}.

\textbf{Reconstruction:} At this stage of the MPQC protocol, each quantum node $i$ acting as a reconstructor $R^i$, collects all the
single-qubit quantum states corresponding to his output logical quantum state $\bar{\bar{\Omega}}^i$ from the other quantum nodes and by
executing the reconstruction phase of the VHSS protocol eventually obtains his single-qubit output quantum state $\omega^i$, see
Table~\ref{table:vhss_protocol}. During the reconstruction phase of the VHSS protocol, the reconstructor $R^i$ creates another public set
of apparent cheaters $\tilde{B}^i$ which records all the arbitrary quantum errors introduced by the cheating quantum nodes at the second
level encoding, and in such a way checks whether each quantum node $j$ is honest or not, i.e., if $|\tilde{B}^i| > t$ is satisfied, the
reconstructor $R^i$ adds quantum node $j$ to the global public set of apparent cheaters $B$. This phase of the MPQC protocol does not
require any additional workspace for the implementation and has a communication complexity of $\mathcal{O}(n^2)$ qubits per quantum node,
which is indeed should be identical to the sharing phase since in terms of the communication complexity they are identical. For the
details see Table~\ref{table:mpqc_protocol}.

% /////////////////////////////////////////////////////////////////////////////////////////////////////////////////////////////////////////
\section{
  \label{sec:security_mpqc_protocol}
  Security proof of the MPQC protocol
}
% /////////////////////////////////////////////////////////////////////////////////////////////////////////////////////////////////////////

In this section, we prove that our MPQC protocol is secure. First, we state the security framework as well as the security definition in
Sec.~\ref{subsec:security_statements}, and second, in Sec.~\ref{subsec:security_proof} we show that the security proof of our MPQC protocol
will be identical to the security proof of the previously suggested MPQC protocol. Finally, to be self-contained, we briefly present the
security proof of our MPQC protocol in Secs.~\ref{subsubsec:ideal_protocol} and \ref{subsubsec:real_protocol}.

% =========================================================================================================================================
\subsection{
  \label{subsec:security_statements}
  Security statements
}
% =========================================================================================================================================

Here we state the security framework and the security definition following Refs.~\cite{beaver_crypto_91_377-391_1992,
micali_crypto_91_392-404_1992,canetti_ieee_42_136-145_2001,unruh_eurocrypt_2010_486-505_2010,lipinska_pra_102_022405_2020}. To prove that
our MPQC protocol is secure we employ the simulator-based security definition, which automatically satisfies requirements of
\textit{correctness}, \textit{soundness}, and \textit{privacy} mentioned in Sec.~\ref{sec:introduction}. The simulator-based security
definition uses two models: the ``real'' model corresponding to the execution of the actual MPQC protocol and the ``ideal'' model where
quantum nodes interact with an oracle that performs the MPQC protocol perfectly and cannot be corrupted by the adversary. In this security
framework, the MPQC protocol is said to be secure if one cannot distinguish a ``real'' execution from an ``ideal'' execution of the MPQC
protocol.

In the ``ideal'' model the honest quantum nodes solely send their input quantum states to the oracle and merely output whatever they
receive from the oracle as their results. On the other hand, cheating quantum nodes are allowed to perform any joint quantum operation on
their input quantum states before sending them to the oracle and also allowed to perform any joint quantum operation on whatever they
receive from the oracle before they output their results. We assume that the cheating quantum nodes are non-adaptively corrupted by an
active adversary $\mathcal{A}$, which can corrupt $t < \frac{n}{4}$ quantum nodes, but otherwise has unlimited computational power, see
Sec.~\ref{subsec:adversary}. Hereafter, an adversary in the ``real'' model will be denoted as $\mathcal{A}_\mathrm{real}$ and an adversary
in the ``ideal'' model will be denoted as $\mathcal{A}_\mathrm{ideal}$.

% |||||||||||||||||||||||||||||||||||||||||||||||||||||||||||||||||||||||||||||||||||||||||||||||||||||||||||||||||||||||||||||||||||||||||
\begin{definition*}
% |||||||||||||||||||||||||||||||||||||||||||||||||||||||||||||||||||||||||||||||||||||||||||||||||||||||||||||||||||||||||||||||||||||||||

The MPQC protocol $\Pi$ is $\epsilon$-secure, if for any input quantum state $\rho$, and for any adversary in the ``real'' model
$\mathcal{A}_\mathrm{real}$, there exists an adversary in the ``ideal'' model $\mathcal{A}_\mathrm{ideal}$, such that the output quantum
state $\omega_\mathrm{real} \coloneqq \Pi_\mathrm{real}(\rho)$ of the ``real'' model is $\epsilon$-close to the output quantum state
$\omega_\mathrm{ideal} \coloneqq \Pi_\mathrm{ideal}(\rho)$ of the ``ideal'' model, i.e.,
% +++++++++++++++++++++++++++++++++++++++++++++++++++++++++++++++++++++++++++++++++++++++++++++++++++++++++++++++++++++++++++++++++++++++++
\begin{align}
\frac{1}{2} \lVert \omega_\mathrm{real} - \omega_\mathrm{ideal} \rVert_1 \leq \epsilon.
\label{eq:epsilon-security}
\end{align}
% +++++++++++++++++++++++++++++++++++++++++++++++++++++++++++++++++++++++++++++++++++++++++++++++++++++++++++++++++++++++++++++++++++++++++

% |||||||||||||||||||||||||||||||||||||||||||||||||||||||||||||||||||||||||||||||||||||||||||||||||||||||||||||||||||||||||||||||||||||||||
\end{definition*}
% |||||||||||||||||||||||||||||||||||||||||||||||||||||||||||||||||||||||||||||||||||||||||||||||||||||||||||||||||||||||||||||||||||||||||

By using the definition of the $\epsilon$-security we can state the security of our MPQC protocol as follows, see
Refs.~\cite{lipinska_pra_102_022405_2020,lipinska_arxiv:2004.10486v2} for details.

% |||||||||||||||||||||||||||||||||||||||||||||||||||||||||||||||||||||||||||||||||||||||||||||||||||||||||||||||||||||||||||||||||||||||||
\begin{theorem}
\label{theorem}
% |||||||||||||||||||||||||||||||||||||||||||||||||||||||||||||||||||||||||||||||||||||||||||||||||||||||||||||||||||||||||||||||||||||||||

The MPQC protocol is $\kappa 2^{-\Omega(r)}$-secure, where $\kappa = n + \#\ancillas + \#H$.

% |||||||||||||||||||||||||||||||||||||||||||||||||||||||||||||||||||||||||||||||||||||||||||||||||||||||||||||||||||||||||||||||||||||||||
\end{theorem}
% |||||||||||||||||||||||||||||||||||||||||||||||||||||||||||||||||||||||||||||||||||||||||||||||||||||||||||||||||||||||||||||||||||||||||

Proof of the security of our MPQC protocol will be almost the same as the security proof of the previously suggested MPQC protocol in
Refs.~\cite{lipinska_pra_102_022405_2020,lipinska_arxiv:2004.10486v2}~\footnote{In the Ref.~\cite{lipinska_arxiv:2004.10486v2} it was shown
that the security proofs of the original version of the MPQC protocol in Ref.~\cite{lipinska_pra_102_022405_2020} and the reconsidered
version of the MPQC protocol in Ref.~\cite{lipinska_arxiv:2004.10486v2} are identical.}, and the only essential difference lies in the type
of the non-transversal quantum gate, namely the $T$ gate is substituted for the $H$ gate, and in the basis of the logical measurement,
namely the normal basis is substituted for the Fourier basis. In particular, in the security proof of the previously suggested MPQC
protocol the ``ideal'' protocol is constructed by using a simulation technique, i.e., for any ``real'' adversary
$\mathcal{A}_\mathrm{real}$ an ``ideal'' adversary $\mathcal{A}_\mathrm{ideal}$ is constructed by saying that an ``ideal'' adversary
$\mathcal{A}_\mathrm{ideal}$ internally simulates the execution of the ``real'' protocol with ``real'' adversary
$\mathcal{A}_\mathrm{real}$. Specifically, one writes the execution of the ``real'' protocol and the ``ideal'' protocol and shows that the
outputs of both protocols are equivalent in case of success of the VHSS protocol. Finally, we note that the security definition employed in
this paper follows the paradigm of sequential composability, see Refs.~\cite{lipinska_pra_102_022405_2020,lipinska_arxiv:2004.10486v2} for
details.

% =========================================================================================================================================
\subsection{
  \label{subsec:security_proof}
  Security proof
}
% =========================================================================================================================================

Here we show that the security proof of our MPQC protocol can be reduced to the security proof of the previously suggested MPQC protocol
presented in Refs.~\cite{lipinska_pra_102_022405_2020,lipinska_arxiv:2004.10486v2}. To achieve that, we borrow statements from the previous
suggestion and restate the lemma with the corresponding proof as will be given below, and in such a way show that there is no difference
between our MPQC protocol and the MPQC protocol in Refs.~\cite{lipinska_pra_102_022405_2020,lipinska_arxiv:2004.10486v2} when the security
proof is the concern. The lemma shows that preparing, sharing, and verifying the input quantum state, then performing logical quantum
circuit $\bar{\bar{\mathcal{U}}}$, and finally reconstructing and measuring the output quantum state is equivalent to preparing the input
quantum state, performing quantum circuit $\mathcal{U}$, and measuring the output quantum state without any encoding. After restating the
lemma, we also restate the property that extends the applicability of the lemma from individual quantum operations to the entire quantum
circuit.

% |||||||||||||||||||||||||||||||||||||||||||||||||||||||||||||||||||||||||||||||||||||||||||||||||||||||||||||||||||||||||||||||||||||||||
\begin{lemma}
\label{lemma}
% |||||||||||||||||||||||||||||||||||||||||||||||||||||||||||||||||||||||||||||||||||||||||||||||||||||||||||||||||||||||||||||||||||||||||

Let us define a public set of apparent cheaters at the end of the computation phase as $B_\mathrm{C}$, such that $|B_\mathrm{C}| \leq t$,
and let us define a set of real cheaters at the end of the computation phase as $A_\mathrm{C}$. Let us also denote the decoding procedure
for the triply-even CSS QECC $\mathcal{C}_\mathrm{TE}$ as $\mathcal{D}$ and the erasure recovery procedure for the triply-even CSS QECC
$\mathcal{C}_\mathrm{TE}$ as $\hat{\mathcal{D}}$. If the global logical quantum state $\bar{\bar{\mathcal{P}}}$ encoded twice by using the
triply-even CSS QECC $\mathcal{C}_\mathrm{TE}$ is decodable, i.e.,
% +++++++++++++++++++++++++++++++++++++++++++++++++++++++++++++++++++++++++++++++++++++++++++++++++++++++++++++++++++++++++++++++++++++++++
\begin{align}
\mathcal{P} = \underset{i \in [n]}{\bigotimes}
\left( \hat{\mathcal{D}}_{\overbar{B_\mathrm{C} \cup A_\mathrm{C}}}
\circ \underset{j \in \overbar{B_\mathrm{C} \cup A_\mathrm{C}}}{\bigotimes} \mathcal{D}_j \right)
(\bar{\bar{\mathcal{P}}}),
\label{eq:state_decodable}
\end{align}
% +++++++++++++++++++++++++++++++++++++++++++++++++++++++++++++++++++++++++++++++++++++++++++++++++++++++++++++++++++++++++++++++++++++++++
then application of a logical quantum operation $\bar{\bar{\mathcal{Q}}}$ to the global logical quantum state $\bar{\bar{\mathcal{P}}}$ is
also decodable, i.e.,
% +++++++++++++++++++++++++++++++++++++++++++++++++++++++++++++++++++++++++++++++++++++++++++++++++++++++++++++++++++++++++++++++++++++++++
\begin{align}
\mathcal{Q}(\mathcal{P}) = \underset{i \in [n]}{\bigotimes}
\left( \hat{\mathcal{D}}_{\overbar{B_\mathrm{C} \cup A_\mathrm{C}}}
\circ \underset{j \in \overbar{B_\mathrm{C} \cup A_\mathrm{C}}}{\bigotimes} \mathcal{D}_j \right)
\left(\bar{\bar{\mathcal{Q}}}(\bar{\bar{\mathcal{P}}})\right),
\label{eq:operation_state_decodable}
\end{align}
% +++++++++++++++++++++++++++++++++++++++++++++++++++++++++++++++++++++++++++++++++++++++++++++++++++++++++++++++++++++++++++++++++++++++++
where the logical quantum operation $\bar{\bar{\mathcal{Q}}}$ may denote:
\begin{itemize}
  \item Logical versions of the transversal quantum gates ($X$ gate, $Z$ gate, $T$ gate, or $\mathrm{C}\text{-}X$ gate) applied to the
  global logical quantum state $\bar{\bar{\mathcal{P}}}$.
  \item Logical version of the non-transversal $H$ gate applied to the global logical quantum state $\bar{\bar{\mathcal{P}}}$ by using the
  gate teleportation protocol.
  \item Logical measurement in the standard or Fourier basis $\bar{\bar{\mathcal{M}}}$ which is implemented by local measurements of the
  single-qubit quantum states, each denoted as $\mathcal{M}$, and the classical communication.
\end{itemize}

% |||||||||||||||||||||||||||||||||||||||||||||||||||||||||||||||||||||||||||||||||||||||||||||||||||||||||||||||||||||||||||||||||||||||||
\end{lemma}
% |||||||||||||||||||||||||||||||||||||||||||||||||||||||||||||||||||||||||||||||||||||||||||||||||||||||||||||||||||||||||||||||||||||||||

% |||||||||||||||||||||||||||||||||||||||||||||||||||||||||||||||||||||||||||||||||||||||||||||||||||||||||||||||||||||||||||||||||||||||||
\begin{proof}
% |||||||||||||||||||||||||||||||||||||||||||||||||||||||||||||||||||||||||||||||||||||||||||||||||||||||||||||||||||||||||||||||||||||||||

The Lemma~\ref{lemma} follows from the fact that to realize a logical quantum operation $\bar{\bar{\mathcal{Q}}}$ it is sufficient to apply
quantum operations $\mathcal{Q}$ honestly on the shares of the quantum nodes in the set $\overbar{B_\mathrm{C} \cup A_\mathrm{C}}$. First,
the application of transversal quantum gates ($X$ gate, $Z$ gate, $T$ gate, and $\mathrm{C}\text{-}X$ gate) on the shares of the quantum
nodes in the set $\overbar{B_\mathrm{C} \cup A_\mathrm{C}}$ indeed realizes the logical versions of these quantum gates ($\bar{\bar{X}}$
gate, $\bar{\bar{Z}}$ gate, $\bar{\bar{T}}$ gate, and $\overbar{\overbar{\mathrm{C}\text{-}X}}$ gate)~\cite{gottesman_pra_57_127_1998}.
Second, in case of CSS QECCs, the logical measurement $\bar{\bar{\mathcal{M}}}$ in the standard or Fourier basis can be implemented
transversally. Third, we implement a non-transversal $H$ gate by combining the transversal quantum gates with the transversally implemented
logical measurement.

% |||||||||||||||||||||||||||||||||||||||||||||||||||||||||||||||||||||||||||||||||||||||||||||||||||||||||||||||||||||||||||||||||||||||||
\end{proof}
% |||||||||||||||||||||||||||||||||||||||||||||||||||||||||||||||||||||||||||||||||||||||||||||||||||||||||||||||||||||||||||||||||||||||||

% |||||||||||||||||||||||||||||||||||||||||||||||||||||||||||||||||||||||||||||||||||||||||||||||||||||||||||||||||||||||||||||||||||||||||
\begin{property}
\label{property}
% |||||||||||||||||||||||||||||||||||||||||||||||||||||||||||||||||||||||||||||||||||||||||||||||||||||||||||||||||||||||||||||||||||||||||

Let us define a quantum circuit as $\mathcal{R}$. Then, Lemma~\ref{lemma} holds even when we replace quantum operation $\mathcal{Q}$ by a
quantum circuit $\mathcal{R}$, i.e.,
% +++++++++++++++++++++++++++++++++++++++++++++++++++++++++++++++++++++++++++++++++++++++++++++++++++++++++++++++++++++++++++++++++++++++++
\begin{align}
\mathcal{R}(\mathcal{P}) = \underset{i \in [n]}{\bigotimes}
\left( \hat{\mathcal{D}}_{\overbar{B_\mathrm{C} \cup A_\mathrm{C}}}
\circ \underset{j \in \overbar{B_\mathrm{C} \cup A_\mathrm{C}}}{\bigotimes} \mathcal{D}_j \right)
\left(\bar{\bar{\mathcal{R}}}(\bar{\bar{\mathcal{P}}})\right).
\label{eq:cirquit_state_decodable}
\end{align}
% +++++++++++++++++++++++++++++++++++++++++++++++++++++++++++++++++++++++++++++++++++++++++++++++++++++++++++++++++++++++++++++++++++++++++

% |||||||||||||||||||||||||||||||||||||||||||||||||||||||||||||||||||||||||||||||||||||||||||||||||||||||||||||||||||||||||||||||||||||||||
\end{property}
% |||||||||||||||||||||||||||||||||||||||||||||||||||||||||||||||||||||||||||||||||||||||||||||||||||||||||||||||||||||||||||||||||||||||||

% |||||||||||||||||||||||||||||||||||||||||||||||||||||||||||||||||||||||||||||||||||||||||||||||||||||||||||||||||||||||||||||||||||||||||
\begin{proof}
% |||||||||||||||||||||||||||||||||||||||||||||||||||||||||||||||||||||||||||||||||||||||||||||||||||||||||||||||||||||||||||||||||||||||||

The Property~\ref{property} immediately follows from the fact that any quantum circuit $\mathcal{R}$ can be decomposed as
$\mathcal{R} = \mathcal{U} \circ \mathcal{M}$, where the quantum circuit $\mathcal{U}$ can be decomposed into the quantum gates chosen so
as to implement the UQC (which are $X$ gate, $Z$ gate, $T$ gate, $\mathrm{C}\text{-}X$ gate, and $H$ gate in our case).

% |||||||||||||||||||||||||||||||||||||||||||||||||||||||||||||||||||||||||||||||||||||||||||||||||||||||||||||||||||||||||||||||||||||||||
\end{proof}
% |||||||||||||||||||||||||||||||||||||||||||||||||||||||||||||||||||||||||||||||||||||||||||||||||||||||||||||||||||||||||||||||||||||||||

With the Property~\ref{property} at hand, it becomes clear that the security proof of our MPQC protocol will be absolutely the same as the
security proof of the previously suggested MPQC protocol presented in Refs.~\cite{lipinska_pra_102_022405_2020,lipinska_arxiv:2004.10486v2}
since the difference between the Property~\ref{property} in our current suggestion and the property in the previous suggestion is reduced
to which quantum gate is implemented by the gate teleportation technique ($H$ gate instead of $T$ gate in our case) and in which basis
logical measurement is performed during the gate teleportation technique (Fourier basis instead of normal basis in our case). Note that in
our MPQC protocol the previously inevitable verification of the ``magic'' state technique, see
Appendix~\ref{sec:magic_state_verification_protocol}, is not necessary at all~\footnote{The same can be said also for the protocol called
``verification of the Clifford stabilized states'' (VCSS), see Appendix~\ref{sec:vcss_summary} and Appendix~\ref{sec:vcss_critics}.}, and
therefore, is out of consideration.

However, solely to be self-contained, we briefly present the security proof of our MPQC protocol, i.e., the proof of Theorem~\ref{theorem}.
Specifically, we follow the security proof of the previously suggested MPQC protocol presented in Refs.~\cite{lipinska_pra_102_022405_2020,
lipinska_arxiv:2004.10486v2}, which was actually inspired by the approach taken in Refs.~\cite{smith_arxiv:quant-ph/0111030,
crepeau_stoc_02_643-652_2002,cramer_damgard_nielsen}. We construct the ``real'' protocol by expressing each quantum operation performed
during the execution of the ``real'' protocol, and consequently the output quantum state of the ``real'' protocol $\omega_\mathrm{real}$
in terms of the general maps, see Sec.~\ref{subsubsec:real_protocol}. Then, the same is done for the ``ideal'' protocol and the output
quantum state of the ``ideal'' protocol $\omega_\mathrm{ideal}$ is also obtained, see Sec.~\ref{subsubsec:ideal_protocol}. Indeed, if these
outputs are compared it becomes clear that they are exponentially close to each other in the security parameter $r$, see
Theorem~\ref{theorem}~\footnote{If we suppose that the VHSS protocol involved in the construction of the MPQC protocol has no any
probability of error, one will actually achieve $\omega_\mathrm{real} = \omega_\mathrm{ideal}$}.

Finally, let us explain where the probability of error in the security statement of the MPQC protocol in Theorem.~\ref{theorem} comes from.
Every verification performed by using the VHSS protocol has the probability of error $2^{-\Omega(r)}$. During the MPQC protocol, the VHSS
protocol is invoked in the following three situations:
\begin{itemize}
  \item When the quantum nodes jointly verify the encoding of each single-qubit input quantum state $\rho^i$ (there are $n$ of them).
  \item When quantum nodes jointly verify and confirm the ancillary logical quantum state $\bar{\bar{\ket{+}}}^i$ necessary for the
  implementation of the non-transversal $H^i$ gate via the gate teleportation technique.
  \item When quantum nodes jointly verify and confirm the ancillary logical quantum state $\bar{\bar{\ket{0}}}^i$ necessary for the
  implementation of the quantum circuit $\mathcal{U}$.
\end{itemize}
If we summarize all the above three cases we obtain the total number of VHSS protocol executions during the MPQC protocol as
$\kappa = n + \#\ancillas + \#H$ and the total probability of error will be $\kappa 2^{-\Omega(r)}$. \qed

% ~~~~~~~~~~~~~~~~~~~~~~~~~~~~~~~~~~~~~~~~~~~~~~~~~~~~~~~~~~~~~~~~~~~~~~~~~~~~~~~~~~~~~~~~~~~~~~~~~~~~~~~~~~~~~~~~~~~~~~~~~~~~~~~~~~~~~~~~~
\subsubsection{
  \label{subsubsec:real_protocol}
  ``Real'' protocol
}
% ~~~~~~~~~~~~~~~~~~~~~~~~~~~~~~~~~~~~~~~~~~~~~~~~~~~~~~~~~~~~~~~~~~~~~~~~~~~~~~~~~~~~~~~~~~~~~~~~~~~~~~~~~~~~~~~~~~~~~~~~~~~~~~~~~~~~~~~~~

Here we construct the ``real'' execution of the MPQC protocol. As will be explained in Sec.~\ref{subsubsec:ideal_protocol}, since in the
``ideal'' protocol the \textit{oracle} receives ``abort'' flag at the end of the computation, in the ``real'' protocol one should also
abort at the end of the computation. However, computation with $|B| > t$ already satisfied may allow cheating quantum nodes to obtain some
information on the inputs of the honest quantum nodes. Therefore, to avoid this situation, the honest quantum nodes replace single-qubit
quantum states in their possession with $\ket{0}$ whenever $|B| > t$ is satisfied, see Ref.~\cite{lipinska_pra_102_022405_2020} for
details.

First of all, let us denote the registers of the honest and cheating quantum nodes in the ``real'' protocol as $H_R$ and $A_R$
respectively. Then, if we denote the general map of the sharing and verification phases as $\mathcal{VS}_{H_R A_R}$, and the input quantum
state of all the quantum nodes as $\rho_{H_R A_R}$, the quantum state after the sharing and the verification will be denoted as
% +++++++++++++++++++++++++++++++++++++++++++++++++++++++++++++++++++++++++++++++++++++++++++++++++++++++++++++++++++++++++++++++++++++++++
\begin{align}
\sigma^\mathcal{VS} = \mathcal{VS}_{H_R A_R} \left( \rho_{H_R A_R} \right).
\label{eq:quantum_state_verification_sharing_real}
\end{align}
% +++++++++++++++++++++++++++++++++++++++++++++++++++++++++++++++++++++++++++++++++++++++++++++++++++++++++++++++++++++++++++++++++++++++++

Then, ``real'' protocol continues to the computation phase. Here, if $|B| \leq t$ is satisfied, all the quantum nodes jointly perform the
logical quantum circuit $\bar{\bar{\mathcal{R}}}_{H_R A_R}$. On the other hand, if $|B| > t$ is satisfied, the honest quantum nodes replace
single-qubit quantum states in their possession with $\ket{0}$ and the cheating quantum nodes perform arbitrary quantum operation
$\mathcal{M}^\prime_{A_R}$ on the shares in their possession. Therefore, the quantum state after the computation will be denoted as
% +++++++++++++++++++++++++++++++++++++++++++++++++++++++++++++++++++++++++++++++++++++++++++++++++++++++++++++++++++++++++++++++++++++++++
\begin{align}
\sigma^\mathcal{R} =
\begin{cases}
\bar{\bar{\mathcal{R}}}_{H_R A_R} \left( \sigma^\mathcal{VS} \right)                                               & |B| \leq t, \\
\mathcal{M}^\prime_{A_R} \circ \mathrm{Tr}_{H_R} \left( \sigma^\mathcal{VS} \right) \otimes \ket{0}\!\bra{0}_{H_R} & |B| > t.
\end{cases}
\label{eq:quantum_state_computation_real}
\end{align}
% +++++++++++++++++++++++++++++++++++++++++++++++++++++++++++++++++++++++++++++++++++++++++++++++++++++++++++++++++++++++++++++++++++++++++

Next, if $|B| \leq t$ is satisfied after the computation phase, the ``real'' protocol continues to the reconstruction phase. The honest
quantum nodes perform the decoding procedure and the erasure recovery procedure, together denoted as $\mathcal{D}_{H_R}$. At the same time,
the cheating quantum nodes perform arbitrary quantum operation $\mathcal{W}_{A_R}$ on the shares in their possession. On the other hand, if
$|B| > t$ is satisfied, the honest quantum nodes output the ``abort'' flag $\ket{\bot}\!\bra{\bot}_{H_R}$. Simultaneously, the cheating
quantum nodes perform arbitrary quantum operation $\mathcal{M}^{\prime\prime}_{A_R}$ on the shares in their possession. Consequently, the
quantum state after the reconstruction will be denoted as
% +++++++++++++++++++++++++++++++++++++++++++++++++++++++++++++++++++++++++++++++++++++++++++++++++++++++++++++++++++++++++++++++++++++++++
\begin{align}
\sigma^\mathcal{D} =
\begin{cases}
\left( \mathcal{D}_{H_R} \otimes \mathcal{W}_{A_R} \right) \left( \sigma^\mathcal{R} \right)
& |B| \leq t, \\
\ket{\bot}\!\bra{\bot}_{H_R} \otimes \mathcal{M}^{\prime\prime}_{A_R} \left[ \mathrm{Tr}_{H_R} \left( \sigma^\mathcal{R} \right) \right]
& |B| > t.
\end{cases}
\label{eq:quantum_state_reconstruction_real}
\end{align}
% +++++++++++++++++++++++++++++++++++++++++++++++++++++++++++++++++++++++++++++++++++++++++++++++++++++++++++++++++++++++++++++++++++++++++

Hereafter, we describe only the case when $|B| \leq t$ is satisfied, since it is enough for our purpose. The case when $|B| > t$ is
satisfied can be found in Ref.~\cite{lipinska_pra_102_022405_2020}. To simplify Eq.~\ref{eq:quantum_state_reconstruction_real}, we
introduce an identity map $\mathbb{I}_{H_R A_R} = \mathcal{D}_{H_R A_R} \circ \mathcal{E}_{H_R A_R}$, where $\mathcal{D}_{H_R A_R}$ and
$\mathcal{E}_{H_R A_R}$ denote the decoding and the encoding procedures respectively. By using this identity map, the output quantum state
$\omega_\mathrm{real}$ of the ``real'' protocol can be written as
% +++++++++++++++++++++++++++++++++++++++++++++++++++++++++++++++++++++++++++++++++++++++++++++++++++++++++++++++++++++++++++++++++++++++++
\begin{align}
\omega_\mathrm{real} = \left( \mathcal{D}_{H_R} \otimes \mathcal{W}_{A_R} \right) \circ \mathcal{E}_{H_R A_R} \circ \mathcal{D}_{H_R A_R}
\left( \sigma^\mathcal{R} \right).
\label{eq:output_quantum_state_success_real}
\end{align}
% +++++++++++++++++++++++++++++++++++++++++++++++++++++++++++++++++++++++++++++++++++++++++++++++++++++++++++++++++++++++++++++++++++++++++

To simplify Eq.~\ref{eq:output_quantum_state_success_real}, we employ the results of the Lemma~\ref{lemma} and the Property~\ref{property},
i.e., that preparing, sharing, and verifying the input quantum state of all the quantum nodes $\rho_{H_R A_R}$, then performing logical
quantum circuit $\bar{\bar{\mathcal{R}}}_{H_R A_R}$, and finally reconstructing and measuring the output quantum state is equivalent to
preparing the input quantum state of all the quantum nodes $\rho_{H_R A_R}$, performing quantum circuit $\mathcal{R}_{H_R A_R}$, and
measuring the output quantum state without any encoding, see Sec.~\ref{subsec:security_proof}. Therefore, the output quantum state
$\omega_\mathrm{real}$ of the ``real'' protocol can be further simplified as
% +++++++++++++++++++++++++++++++++++++++++++++++++++++++++++++++++++++++++++++++++++++++++++++++++++++++++++++++++++++++++++++++++++++++++
\begin{align}
\omega_\mathrm{real} = \left( \mathcal{D}_{H_R} \otimes \mathcal{W}_{A_R} \right) \circ \mathcal{E}_{H_R A_R} \circ \mathcal{R}_{H_R A_R}
\left( \rho_{H_R A_R} \right).
\label{eq:output_quantum_state_success_simplified_real}
\end{align}
% +++++++++++++++++++++++++++++++++++++++++++++++++++++++++++++++++++++++++++++++++++++++++++++++++++++++++++++++++++++++++++++++++++++++++

% ~~~~~~~~~~~~~~~~~~~~~~~~~~~~~~~~~~~~~~~~~~~~~~~~~~~~~~~~~~~~~~~~~~~~~~~~~~~~~~~~~~~~~~~~~~~~~~~~~~~~~~~~~~~~~~~~~~~~~~~~~~~~~~~~~~~~~~~~~
\subsubsection{
  \label{subsubsec:ideal_protocol}
  ``Ideal'' protocol
}
% ~~~~~~~~~~~~~~~~~~~~~~~~~~~~~~~~~~~~~~~~~~~~~~~~~~~~~~~~~~~~~~~~~~~~~~~~~~~~~~~~~~~~~~~~~~~~~~~~~~~~~~~~~~~~~~~~~~~~~~~~~~~~~~~~~~~~~~~~~

% -----------------------------------------------------------------------------------------------------------------------------------------
\begin{figure}[htb]
  \begin{center}
% !!!!!!!!!!!!!!!!!!!!!!!!!!!!!!!!!!!!!!!!!!!!!!!!!!!!!!!!!!!!!!!!!!!!!!!!!!!!!!!!!!!!!!!!!!!!!!!!!!!!!!!!!!!!!!!!!!!!!!!!!!!!!!!!!!!!!!!!!
    \includegraphics[width=\columnwidth]{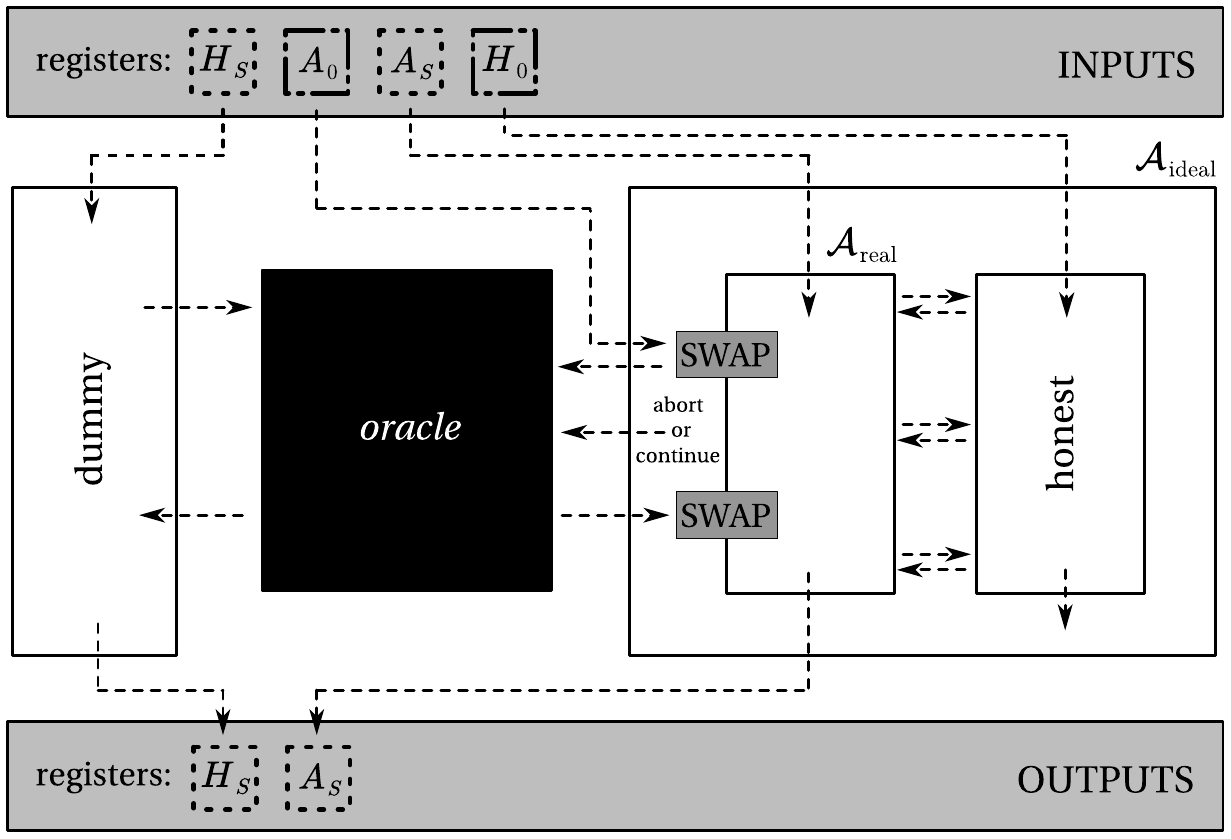}
% !!!!!!!!!!!!!!!!!!!!!!!!!!!!!!!!!!!!!!!!!!!!!!!!!!!!!!!!!!!!!!!!!!!!!!!!!!!!!!!!!!!!!!!!!!!!!!!!!!!!!!!!!!!!!!!!!!!!!!!!!!!!!!!!!!!!!!!!!
    \caption{
      Schematic picture of the simulator-based security proof of the MPQC protocol. The ``ideal'' execution of the MPQC protocol requires
      following four types of quantum registers: registers of the simulated honest quantum nodes $H_0$, registers of the simulated cheating
      quantum nodes $A_0$, ``dummy'' input registers of the honest quantum nodes in the simulation $H_S$, and input registers of the
      cheating quantum nodes in the simulation $A_S$. Also, ``ideal'' protocol requires a classical flag to decide whether to abort the
      MPQC or not, which is denoted as ``abort'' or ``continue''.
    }
% !!!!!!!!!!!!!!!!!!!!!!!!!!!!!!!!!!!!!!!!!!!!!!!!!!!!!!!!!!!!!!!!!!!!!!!!!!!!!!!!!!!!!!!!!!!!!!!!!!!!!!!!!!!!!!!!!!!!!!!!!!!!!!!!!!!!!!!!!
    \label{fig:security_proof}
% !!!!!!!!!!!!!!!!!!!!!!!!!!!!!!!!!!!!!!!!!!!!!!!!!!!!!!!!!!!!!!!!!!!!!!!!!!!!!!!!!!!!!!!!!!!!!!!!!!!!!!!!!!!!!!!!!!!!!!!!!!!!!!!!!!!!!!!!!
  \end{center}
\end{figure}
% -----------------------------------------------------------------------------------------------------------------------------------------

Next we construct the ``ideal'' execution of the MPQC protocol. The adversary in the ``ideal'' protocol $\mathcal{A}_\mathrm{ideal}$ will
internally simulate the ``real'' protocol with the ``real'' adversary $\mathcal{A}_\mathrm{real}$. Here, the simulated honest quantum nodes
will interact with the simulated cheating quantum nodes controlled by the ``real'' adversary $\mathcal{A}_\mathrm{real}$, see
Fig.~\ref{fig:security_proof}. In the ``ideal'' protocol, the ``ideal'' adversary $\mathcal{A}_\mathrm{ideal}$ and the honest quantum nodes
interact with an \textit{oracle} that perfectly realizes the MPQC protocol and cannot be corrupted. As an input, the \textit{oracle}
requires ``dummy'' quantum registers of the honest quantum nodes in the simulation $H_S$, quantum registers of the cheating quantum nodes
in the simulation $A_S$, and a classical flag which indicates whether the \textit{oracle} should abort the ``ideal'' protocol or not.

If we denote the input quantum state of all the quantum nodes in the simulation as $\rho_{H_S A_S}$ the entire input into the ``ideal''
protocol will be $\rho_{H_S A_S} \otimes \ket{0}\!\bra{0}_{H_0 A_0}$. Furthermore, if we denote the general map of the sharing and
verification phases as $\mathcal{VS}_{H_0 A_S}$, the quantum state after the sharing and the verification will be denoted as
% +++++++++++++++++++++++++++++++++++++++++++++++++++++++++++++++++++++++++++++++++++++++++++++++++++++++++++++++++++++++++++++++++++++++++
\begin{align}
\sigma^\mathcal{VS} = \mathcal{VS}_{H_0 A_S} \left( \rho_{H_S A_S} \otimes \ket{0}\!\bra{0}_{H_0 A_0} \right).
\label{eq:quantum_state_verification_sharing_ideal}
\end{align}
% +++++++++++++++++++++++++++++++++++++++++++++++++++++++++++++++++++++++++++++++++++++++++++++++++++++++++++++++++++++++++++++++++++++++++

Before the ``ideal'' protocol continues to the computation phase, the ``ideal'' adversary $\mathcal{A}_\mathrm{ideal}$ performs an
encoding procedure $\mathcal{E}_{A_0}$ and subsequently applies a $\mathrm{SWAP}$ gate between the registers $A_0$ and $A_S$.
\begin{enumerate}[label=(\alph*)]
  \item Here, if $|B| \leq t$ is satisfied, the ``ideal'' adversary $\mathcal{A}_\mathrm{ideal}$ performs an erasure recovery procedure
  twice, which we denoted as $\mathcal{D}_{A_0}$, on the registers of the quantum nodes not in the public set of apparent cheaters $B$ and
  sends register $A_0$ to the \textit{oracle}.
  \item On the other hand, if $|B| > t$ is satisfied, the simulated honest quantum nodes replace single-qubit quantum states in their
  possession with $\ket{0}$, while the ``ideal'' adversary $\mathcal{A}_\mathrm{ideal}$ sends single-qubit quantum states $\ket{0}$, as
  inputs of the simulated cheating quantum nodes, to the \textit{oracle}. Also, simulated cheating quantum nodes perform arbitrary quantum
  operation $\mathcal{M}^\prime_{A_S}$ on the shares in their possession.
\end{enumerate}
Therefore, the quantum state after the first interaction with the \textit{oracle} can be written as
% +++++++++++++++++++++++++++++++++++++++++++++++++++++++++++++++++++++++++++++++++++++++++++++++++++++++++++++++++++++++++++++++++++++++++
\begin{align}
\sigma^{or.^1} =
\begin{cases}
\mathcal{D}_{A_0} \circ \mathrm{SWAP}_{A_0 A_S} \circ \mathcal{E}_{A_0} \left( \sigma^\mathcal{VS} \right)           & |B| \leq t, \\
\mathcal{M}^\prime_{A_S} \otimes \mathrm{Tr}_{H_0} \left( \sigma^\mathcal{VS} \right) \otimes \ket{0}\!\bra{0}_{H_0} & |B| > t.
\end{cases}
\label{eq:first_communication_with_oracle_ideal}
\end{align}
% +++++++++++++++++++++++++++++++++++++++++++++++++++++++++++++++++++++++++++++++++++++++++++++++++++++++++++++++++++++++++++++++++++++++++

Then, the ``ideal'' adversary $\mathcal{A}_\mathrm{ideal}$ proceeds to the computation phase on the registers $H_0$ and $A_S$. Meanwhile,
the \textit{oracle} performs the ideal quantum circuit $\mathcal{R}^\mathrm{ideal}_{H_S A_0}$. Therefore, the quantum state after these
quantum operations can be written as
% +++++++++++++++++++++++++++++++++++++++++++++++++++++++++++++++++++++++++++++++++++++++++++++++++++++++++++++++++++++++++++++++++++++++++
\begin{align}
\sigma^\mathcal{R} =
\begin{cases}
\left( \mathcal{R}^\mathrm{ideal}_{H_S A_0} \otimes \bar{\bar{\mathcal{R}}}_{H_0 A_S} \right)
\left( \sigma^{or.^1} \right) & |B| \leq t, \\
\left( \mathcal{R}^\mathrm{ideal}_{H_S A_0} \otimes \bar{\bar{\mathcal{R}}}_{H_0 A_S} \right)
\left( \sigma^{or.^1} \right) & |B| > t.
\end{cases}
\label{eq:quantum_state_computation_ideal}
\end{align}
% +++++++++++++++++++++++++++++++++++++++++++++++++++++++++++++++++++++++++++++++++++++++++++++++++++++++++++++++++++++++++++++++++++++++++

Next, depending on the number of apparent cheaters, the ``ideal'' adversary and the \textit{oracle} will behave in the following two ways:
\begin{enumerate}[label=(\alph*)]
  \item If $|B| \leq t$ is satisfied, the ``ideal'' adversary $\mathcal{A}_\mathrm{ideal}$ sends the flag ``continue'' to the
  \textit{oracle} and the \textit{oracle} outputs $\ket{\bot}\!\bra{\bot}$.
  \item On the other hand, if $|B| > t$ is satisfied, the ``ideal'' adversary $\mathcal{A}_\mathrm{ideal}$ sends the flag ``abort'' to the
  \textit{oracle} and the \textit{oracle} outputs the result of the ideal quantum circuit $\mathcal{R}^\mathrm{ideal}_{H_S A_0}$
  evaluation.
\end{enumerate}

After that, the honest quantum nodes in the simulation output whatever they receive from the \textit{oracle} as their results. On the other
hand, after receiving the output of the \textit{oracle} the ``ideal'' adversary $\mathcal{A}_\mathrm{ideal}$ does the following:
\begin{enumerate}[label=(\alph*)]
  \item If $|B| \leq t$ is satisfied, the ``ideal'' adversary $\mathcal{A}_\mathrm{ideal}$ performs an encoding procedure twice on the
  registers $A_0$, which we denoted as $\mathcal{E}_{A_0}$. Finally, the ``ideal'' adversary $\mathcal{A}_\mathrm{ideal}$ applies a
  $\mathrm{SWAP}$ gate between the registers $A_S$ and $A_0$.
  \item If $|B| > t$ is satisfied, the simulated ``real'' protocol aborts and the ``ideal'' adversary $\mathcal{A}_\mathrm{ideal}$ outputs
  the result of the ``real'' adversary $\mathcal{A}_\mathrm{real}$. Finally, simulated cheating quantum nodes perform arbitrary quantum
  operation $\mathcal{M}^{\prime\prime}_{A_S}$ on the shares in their possession.
\end{enumerate}
The quantum state after the second interaction with the \textit{oracle} can be written as
% +++++++++++++++++++++++++++++++++++++++++++++++++++++++++++++++++++++++++++++++++++++++++++++++++++++++++++++++++++++++++++++++++++++++++
\begin{align}
\sigma^{or.^2} =
\begin{cases}
\mathrm{SWAP}_{A_0 A_S} \circ \mathcal{E}_{A_0} \left( \sigma^\mathcal{R} \right) & |B| \leq t, \\
\ket{\bot}\!\bra{\bot}_{H_S A_0} \otimes \mathrm{Tr}_{H_S A_0}
\left[ \mathcal{M}^{\prime\prime}_{A_S} \left( \sigma^\mathcal{R} \right) \right] & |B| > t.
\end{cases}
\label{eq:second_communication_with_oracle_ideal}
\end{align}
% +++++++++++++++++++++++++++++++++++++++++++++++++++++++++++++++++++++++++++++++++++++++++++++++++++++++++++++++++++++++++++++++++++++++++

Hereafter, we describe only the case when $|B| \leq t$ is satisfied, since it is enough for our purpose. The case when $|B| > t$ is
satisfied can be found in Ref.~\cite{lipinska_pra_102_022405_2020} as well. To simplify
Eq.~\ref{eq:second_communication_with_oracle_ideal} we employ the identity which holds for any quantum operation $Q_{H_0 A_0 H_S A_S}$ and
can be written as $\mathrm{SWAP}_{A_0 A_S} \circ Q_{H_0 A_0 H_S A_S} \circ \mathrm{SWAP}_{A_0 A_S} = Q_{H_0 H_S A_0 A_S}$. By using this
identity, as well as Eq.~\ref{eq:quantum_state_verification_sharing_ideal}, the simplified quantum state after the second interaction with
the oracle and in the case when $|B| \leq t$ is satisfied can be written as
% +++++++++++++++++++++++++++++++++++++++++++++++++++++++++++++++++++++++++++++++++++++++++++++++++++++++++++++++++++++++++++++++++++++++++
\begin{align}
\sigma^\mathrm{simp.} &= \left( \mathcal{E}_{A_S} \circ \mathcal{R}^\mathrm{ideal}_{H_S A_S} \circ \mathcal{D}_{A_S} \right) \otimes
\left( \bar{\bar{R}}_{H_0 A_0} \circ \mathcal{E}_{A_0} \right) \left( \sigma^\mathcal{VS} \right) \nonumber \\
&= \left(
\mathcal{E}_{A_S} \circ \mathcal{R}^\mathrm{ideal}_{H_S A_S} \circ \mathcal{D}_{A_S} \circ \mathcal{VS}_{A_S} \left( \rho_{H_S A_S} \right)
\right) \nonumber \\
&\otimes \left(
\bar{\bar{R}}_{H_0 A_0} \circ \mathcal{E}_{A_0} \circ \mathcal{VS}_{A_0} \left( \ket{\bot}\!\bra{\bot}_{H_0 A_0} \right)
\right).
\label{eq:swap_swap_simplified_ideal}
\end{align}
% +++++++++++++++++++++++++++++++++++++++++++++++++++++++++++++++++++++++++++++++++++++++++++++++++++++++++++++++++++++++++++++++++++++++++
Note that the simplification in Eq.~\ref{eq:swap_swap_simplified_ideal} means that the composition of two $\mathrm{SWAP}$ gates between the
registers $A_0$ and $A_S$ with the ideal quantum circuit $\mathcal{R}^\mathrm{ideal}_{H_S A_0}$ performed by the \textit{oracle} is
equivalent to the evaluation of the ideal quantum circuit $\mathcal{R}^\mathrm{ideal}_{H_S A_S}$ by the \textit{oracle}.

Finally, the ``ideal'' adversary $\mathcal{A}_\mathrm{ideal}$ proceeds to the reconstruction phase, in which the simulated honest quantum
nodes perform the decoding procedure and the erasure recovery procedure, together denoted as $\mathcal{D}_{H_0}$. Simultaneously, the
simulated cheating quantum nodes perform arbitrary quantum operation $\mathcal{W}_{A_S}$ on the shares in their possession and the
``ideal'' adversary $\mathcal{A}_\mathrm{ideal}$ outputs the result of the ``real'' adversary $\mathcal{A}_\mathrm{real}$. Therefore, the
output quantum state $\omega_\mathrm{ideal}$ of the ``ideal'' protocol can be written as
% +++++++++++++++++++++++++++++++++++++++++++++++++++++++++++++++++++++++++++++++++++++++++++++++++++++++++++++++++++++++++++++++++++++++++
\begin{align}
\omega_\mathrm{ideal} = \mathrm{Tr}_{H_0 A_0}
\left[ \mathcal{D}_{H_0} \otimes \mathcal{W}_{A_S} \left( \sigma^\mathrm{simp.} \right) \right],
\label{eq:output_quantum_state_success_ideal}
\end{align}
% +++++++++++++++++++++++++++++++++++++++++++++++++++++++++++++++++++++++++++++++++++++++++++++++++++++++++++++++++++++++++++++++++++++++++
and, if we employ the identity maps $\mathbb{I}_{A_S} = \mathcal{D}_{A_S} \circ \mathcal{VS}_{A_S}$ and
$\mathbb{I}_{H_S} = \mathcal{D}_{H_S} \circ \mathcal{E}_{H_S}$, Eq.~\ref{eq:output_quantum_state_success_ideal} can be further simplified
as
% +++++++++++++++++++++++++++++++++++++++++++++++++++++++++++++++++++++++++++++++++++++++++++++++++++++++++++++++++++++++++++++++++++++++++
\begin{align}
\omega_\mathrm{ideal} = \left( \mathcal{D}_{H_S} \otimes \mathcal{W}_{A_S} \right) \circ
\mathcal{E}_{H_S A_S} \circ \mathcal{R}^\mathrm{ideal}_{H_S A_S} \left( \rho_{H_S A_S} \right).
\label{eq:output_quantum_state_success_simplified_ideal}
\end{align}
% +++++++++++++++++++++++++++++++++++++++++++++++++++++++++++++++++++++++++++++++++++++++++++++++++++++++++++++++++++++++++++++++++++++++++

% /////////////////////////////////////////////////////////////////////////////////////////////////////////////////////////////////////////
\section{
  \label{sec:summary}
  Summary
}
% /////////////////////////////////////////////////////////////////////////////////////////////////////////////////////////////////////////

To summarize, in this paper we suggested an MPQC protocol built upon a technique of quantum error correction and in particular constructed
on the basis of the triply-even CSS QECCs. With the triply-even CSS QECCs at hand, once we decide on the $X$ gate, $Z$ gate, $T$ gate,
$\mathrm{C}\text{-}X$ gate, and $H$ gate as our universal set of quantum gates, since all the transversal quantum gates can be implemented
trivially, the task of the UQC realization in the MPQC protocol reduces to the implementation of the non-transversal $H$ gate, which can be
easily addressed by the gate teleportation technique. Importantly, this technique requires a logical ``plus'' state as an ancillary quantum
state, which preparation, verification, and confirmation can be accomplished by using the VHSS protocol only. In contrast, the previously
suggested MPQC protocol was constructed on the basis of the self-dual CSS QECCs, in which case, the task of the UQC realization cannot be
attained without the implementation of the non-transversal $T$ gate and the gate teleportation technique comes to aid again. Crucially, the
implementation of the non-transversal $T$ gate with the gate teleportation technique requires a logical ``magic'' state as an ancillary
quantum state, which preparation, verification, and confirmation can be accomplished only by using a combination of the two sub-protocols:
the VHSS protocol and the protocol verifying the ``magic'' state, the latter of which is implemented by a statistical testing of the
randomly selected ``magic'' states with their subsequent distillation~\footnote{The original version of the protocol verifying the
``magic'' state employed a non-transversal $\mathrm{C}\text{-}XP^\dag$ gate potentially leading to a failure of the entire MPQC protocol.}.
Therefore, our decision on the triply-even CSS QECCs allows us to avoid execution of the resource-intensive protocol verifying the
``magic'' state and consequently reduce our demand for the workspace per quantum node from $n^2 + \Theta(r)n$ qubits in the previous
suggestion to $n^2 + 3n$ qubits in our case, where $n$ is the number of quantum nodes participating in the MPQC protocol and $r$ is the
security parameter. Besides, since every extra qubit reduces the credibility of physical devices, our suggestion makes the MPQC protocol
more accessible for the near-future technology.

% /////////////////////////////////////////////////////////////////////////////////////////////////////////////////////////////////////////
% Acknowledgments (begin)
% /////////////////////////////////////////////////////////////////////////////////////////////////////////////////////////////////////////

\begin{acknowledgments}
The authors would like to thank Suguru Endo, Kaoru Yamamoto, Yuuki Tokunaga, and especially Yasunari Suzuki, for fruitful discussions on
the techniques of quantum error correction. The authors also acknowledge Akinori Hosoyamada for insightful comments on the techniques of
classical cryptography.
\end{acknowledgments}

% /////////////////////////////////////////////////////////////////////////////////////////////////////////////////////////////////////////
% Acknowledgments (end)
% /////////////////////////////////////////////////////////////////////////////////////////////////////////////////////////////////////////

\bibliography{main_10}

\appendix

% /////////////////////////////////////////////////////////////////////////////////////////////////////////////////////////////////////////
\section{
  \label{sec:vcss_summary}
  Summary of the VCSS protocol
}
% /////////////////////////////////////////////////////////////////////////////////////////////////////////////////////////////////////////

Here we briefly describe the VCSS protocol which is necessary for the verification of the ancillary logical ``magic'' state
$\bar{\bar{\ket{m}}}^i$, i.e., whether it is certainly a logical ``magic'' state or not, in case of the original version of the MPQC
protocol based on self-dual CSS QECCs~\cite{lipinska_pra_102_022405_2020}. The idea of the VCSS protocol construction is inspired by the
procedure of the stabilizer measurement in the technique of quantum error correction. To begin with, consider $XP^\dag$ gate, and the
``magic'' state $\ket{m} = \frac{1}{\sqrt{2}}(\ket{0} + e^{i\pi/4}\ket{1})$ which is a $+1$ eigenstate of the $XP^\dag$ gate (see
Sec.~\ref{sec:vcss_critics} for the criticism towards this claim). Then the equation
$\mathrm{C}\text{-}XP^\dag(\ket{+}\ket{m}) = \ket{+}\ket{m}$ holds, where $\mathrm{C}\text{-}XP^\dag$ gate is applied between the
single-qubit quantum state $\ket{+}$ acting as the control quantum state and the ancillary ``magic'' state $\ket{m}$ acting as the target
quantum state. This insight suggests on how to implement the verification of the $\ket{m}$. If the target quantum state was $\ket{m}$, then
after applying $\mathrm{C}\text{-}XP^\dag$ gate one will always measure the control quantum state in the $\ket{+}$. On the other hand, if
the target quantum state was not $\ket{m}$ and one measures the control quantum state in the $\ket{+}$, then one has projected the target
quantum state onto the $\ket{m}$.

The above procedure can be adapted to confirm that the quantum state $\bar{\bar{\ket{m}}}^i$ in the possession of the quantum nodes is for
sure an anticipated ancillary logical ``magic'' state. First, by using the VHSS protocol, quantum nodes jointly verify and confirm the
logical quantum state $_v\!\bar{\bar{\ket{0}}}^i$, see Sec.~\ref{subsec:vhss_protocol}. Second, by using the VHSS protocol one more time
quantum nodes jointly verify that the quantum state $\bar{\bar{\ket{m}}}^i$ in their possession is for sure a valid logical quantum state
encoded by a self-dual CSS QECC, see Sec.~\ref{subsec:vhss_protocol}. Next, to obtain the logical quantum state $\bar{\bar{\ket{+}}}^i$
quantum nodes transversally apply $\bar{\bar{H}}^i$ gate~\footnote{For the self-dual CSS QECCs $H$ gate is transversal.} to the logical
quantum state $_v\!\bar{\bar{\ket{0}}}^i$, and subsequently apply $\overbar{\overbar{\mathrm{C}\text{-}XP^\dag}}^i$ gate to their shares,
taking shares of the logical quantum state $\bar{\bar{\ket{+}}}^i$ as the control shares and shares of the logical quantum state
$\bar{\bar{\ket{m}}}^i$ as the target shares. Then, quantum nodes one more time transversally apply $\bar{\bar{H}}^i$ gate to the control
shares and logically measure them in the standard basis. Finally, quantum nodes decode the result of the logical measurement twice and
publicly check whether their twice decoded result corresponds to $\ket{0}^i$. In parallel, quantum nodes update a public set of apparent
cheaters $B$. Note that the above procedure works if and only if the $\mathrm{C}\text{-}XP^\dag$ gate is transversal for a self-dual CSS
QECC (see Sec.~\ref{sec:vcss_critics} for the criticism towards this claim).

Execution of the VCSS protocol requires a workspace of $4n$ qubits per quantum node. First, verification of the logical quantum state
$\bar{\bar{\ket{m}}}^i$ requires a workspace of $3n$ qubits per quantum node. Second, after the verification of the logical quantum state
$\bar{\bar{\ket{m}}}^i$, each quantum node requires a workspace of $n$ qubits for holding this logical quantum state and in addition, uses
extra workspace of $3n$ qubits for verification of the logical quantum state $_v\!\bar{\bar{\ket{0}}}^i$. Thus in total, each quantum node
requires a workspace of $4n$ qubits for the execution of the VCSS protocol. The communication complexity of the VCSS protocol is the same
as of the VHSS protocol, i.e., $\mathcal{O}(nr^2)$ qubits per quantum node.

% /////////////////////////////////////////////////////////////////////////////////////////////////////////////////////////////////////////
\section{
  \label{sec:vcss_critics}
  Criticism towards the VCSS protocol
}
% /////////////////////////////////////////////////////////////////////////////////////////////////////////////////////////////////////////

Here we describe two problems underlying the implementation of the VCSS protocol suggested in Ref.~\cite{lipinska_pra_102_022405_2020}.
Authors of the Ref.~\cite{lipinska_pra_102_022405_2020} claim that the ``magic'' state $\ket{m}$ is a $+1$ eigenstate of the $XP^\dag$ gate
while in reality, it is a $e^{i7\pi/4}$ eigenstate, see Ref.~\cite{holting} for calculations. This fact alone may cause some problems when
applying $\mathrm{C}\text{-}XP^\dag$ gate during the execution of the VCSS protocol, but the issue can be easily fixed by applying
$\mathrm{C}\text{-}e^{i\pi/4}XP^\dag$ gate instead, since $\ket{m}$ is indeed a $+1$ eigenstate of the $e^{i\pi/4}XP^\dag$ gate.
Furthermore, authors of the Ref.~\cite{lipinska_pra_102_022405_2020} claim that the VCSS protocol works as long as the
$\mathrm{C}\text{-}XP^\dag$ gate can be implemented transversally. However, the $\mathrm{C}\text{-}XP^\dag$ gate is not transversal for the
self-dual CSS QECCs. First of all, it is known to be impossible to implement an entire universal set of quantum gates transversally for any
QECC~\cite{eastin_prl_102_110502_2009}. Therefore, the $\mathrm{C}\text{-}XP^\dag$ gate needs to be a Clifford gate. Actually,
$\mathrm{C}\text{-}XP^\dag$ gate can be easily decomposed as
$\mathrm{C}\text{-}XP^\dag = \mathrm{C}\text{-}X \circ \mathrm{C}\text{-}P^\dag$, which implies that the $\mathrm{C}\text{-}P^\dag$ gate is
a Clifford gate. But the $\mathrm{C}\text{-}P^\dag$ gate is obviously not a Clifford gate and we attain a contradiction.

% /////////////////////////////////////////////////////////////////////////////////////////////////////////////////////////////////////////
\section{
  \label{sec:magic_state_verification_protocol}
  Outline of the ``magic'' state verification protocol
}
% /////////////////////////////////////////////////////////////////////////////////////////////////////////////////////////////////////////

Here we briefly describe the protocol necessary for the verification of the ancillary logical ``magic'' state $\bar{\bar{\ket{m}}}^i$,
i.e., whether it is certainly logical ``magic'' state or not, in case of the reconsidered version of the MPQC protocol suggested in
Ref.~\cite{lipinska_arxiv:2004.10486v2}. The protocol we describe here circumvents the questionable applicability of the VCSS protocol
suggested in the original version of the MPQC protocol in Ref.~\cite{lipinska_pra_102_022405_2020}. The ambiguity in the VCSS protocol
comes from the engagement of the $\mathrm{C}\text{-}XP^\dag$ gate, which is non-transversal in case of the self-dual CSS QECCs.

The protocol verifying the ancillary logical ``magic'' state $\bar{\bar{\ket{m}}}^i$ relies on a statistical testing of the randomly
selected ``magic'' states $\ket{m}^i$, with the subsequent distillation of the logical ``magic'' states $\bar{\bar{\ket{m}}}^i$ via the
distributed version of the $15$-to-$1$ magic state distillation protocol~\cite{bravyi_pra_71_022316_2005}. This approach increases the
workspace required for the implementation of the MPQC protocol from $n^2 + 4n$ qubits per quantum node in the original
suggestion~\cite{lipinska_pra_102_022405_2020} to $n^2 + \Theta(r)n$ qubits per quantum node in the reconsidered
suggestion~\cite{lipinska_arxiv:2004.10486v2}, where $n$ is the number of the quantum nodes and $r$ is the security parameter. Fortunately,
the security proof does not change between the two versions of the MPQC protocol.

In short, the verification of the ``magic'' state technique is performed as follows~\cite{lipinska_arxiv:2004.10486v2}. First of all,
quantum nodes jointly prepare $M$ copies of the verified by the VHSS protocol logical ``magic'' state $\bar{\bar{\ket{m}}}^i$, see
Sec.~\ref{subsec:vhss_protocol}. Next, by using the public source of randomness, quantum nodes jointly select $k$ out of $M$ copies, and to
perform the statistical testing of the randomly selected ``magic'' states, randomly ascribe some quantum node $j$ to each selected copy.
After that, each quantum node $j$ collects all the single-qubit quantum states corresponding to the copy ascribed to him and by decoding it
twice reconstructs a ``magic'' state $\ket{m}^i$, see Sec.~\ref{subsec:vhss_protocol}. Then, each quantum node $j$ measures the
reconstructed ``magic'' state in the $\{\ket{m}, \ket{m^\perp}\}$ basis and if all the measurement results correspond to the ``magic''
state $\ket{m}^i$ then the quantum nodes can be sure that the remaining $M - k$ copies of the logical quantum state $\bar{\bar{\ket{m}}}^i$
in their possession is for sure an anticipated logical ``magic'' states with high probability. In parallel, quantum nodes update a public
set of apparent cheaters $B$. After that, the dephasing procedure is performed, where by using the public source of randomness, quantum
nodes randomly apply the $\overbar{\overbar{PX}}^i$ gate to each of the remaining $M - k$ copies of the logical quantum state
$\bar{\bar{\ket{m}}}^i$ in such a way bringing them into the form diagonal in the $\{\ket{m}, \ket{m^\perp}\}$ basis, and subsequently
randomly permute these $M - k$ copies of the logical quantum state $\bar{\bar{\ket{m}}}^i$. Finally, quantum nodes jointly perform the
distillation of the logical ``magic'' state $\bar{\bar{\ket{m}}}^i$ by using the distributed version of the $15$-to-$1$ magic state
distillation protocol~\cite{bravyi_pra_71_022316_2005} which can be implemented by combining the transversal quantum gates with the
transversally implemented logical measurements in case of the self-dual CSS QECCs.

Execution of the above protocol requires a workspace of $(M + 2)n = \Theta(r)n$ qubits per quantum node, since during the verification of
the ``magic'' state quantum nodes should jointly prepare $M$ copies of the verified by the VHSS protocol logical ``magic'' state
$\bar{\bar{\ket{m}}}^i$. The communication complexity of the above protocol is $(Mr^2 + k)n = \mathcal{O}\big(\Theta(r)nr^2\big)$ qubits
per quantum node since in addition to the $M$ executions of the VHSS protocol the verification of the ``magic'' state technique requires a
collection of the $k$ randomly selected copies of the logical ``magic'' state $\bar{\bar{\ket{m}}}^i$.

% /////////////////////////////////////////////////////////////////////////////////////////////////////////////////////////////////////////
\section{
  \label{sec:quantum_gates_definitions}
  Definitions of the quantum gates
}
% /////////////////////////////////////////////////////////////////////////////////////////////////////////////////////////////////////////

We define single-qubit quantum gates used throughout the manuscript, i.e., $X$ gate, $Y$ gate, $Z$ gate, $H$ gate, $P$ gate, and $T$ gate
in a matrix form as follows:
% +++++++++++++++++++++++++++++++++++++++++++++++++++++++++++++++++++++++++++++++++++++++++++++++++++++++++++++++++++++++++++++++++++++++++
\begin{align}
X &= \begin{pmatrix}
     0 &  1 \\
     1 &  0
\end{pmatrix},
&
Y &= \begin{pmatrix}
     0 & -i \\
     i &  0
\end{pmatrix},
&
Z &= \begin{pmatrix}
     1 &  0 \\
     0 & -1
\end{pmatrix}, \nonumber \\ \nonumber \\
H &= \frac{1}{\sqrt{2}}\begin{pmatrix}
     1 &  1 \\
     1 & -1
\end{pmatrix},
&
P &= \begin{pmatrix}
     1 &  0 \\
     0 &  i
\end{pmatrix},
&
T &= \begin{pmatrix}
     1 &  0 \\
     0 &  e^{i\pi/4}
\end{pmatrix}. \nonumber \\ \nonumber \\
\end{align}
% +++++++++++++++++++++++++++++++++++++++++++++++++++++++++++++++++++++++++++++++++++++++++++++++++++++++++++++++++++++++++++++++++++++++++

As well, we define two-qubit quantum gates used throughout the manuscript, i.e., $\mathrm{C}\text{-}X$ gate, and $\mathrm{C}\text{-}P^\dag$
gate as follows:
% +++++++++++++++++++++++++++++++++++++++++++++++++++++++++++++++++++++++++++++++++++++++++++++++++++++++++++++++++++++++++++++++++++++++++
\begin{align}
\mathrm{C}\text{-}X &= \begin{pmatrix}
     1 &  0 &  0 &  0 \\
     0 &  1 &  0 &  0 \\
     0 &  0 &  0 &  1 \\
     0 &  0 &  1 &  0
\end{pmatrix},
&
\mathrm{C}\text{-}P^\dag &= \begin{pmatrix}
     1 &  0 &  0 &  0 \\
     0 &  1 &  0 &  0 \\
     0 &  0 &  1 &  0 \\
     0 &  0 &  0 &  i
\end{pmatrix}. \nonumber \\ \nonumber \\
\end{align}
% +++++++++++++++++++++++++++++++++++++++++++++++++++++++++++++++++++++++++++++++++++++++++++++++++++++++++++++++++++++++++++++++++++++++++

% /////////////////////////////////////////////////////////////////////////////////////////////////////////////////////////////////////////
\section{
  \label{sec:h_gate_teleportation}
  Teleportation of the $H$ gate
}
% /////////////////////////////////////////////////////////////////////////////////////////////////////////////////////////////////////////

% -----------------------------------------------------------------------------------------------------------------------------------------
\begin{figure}[t]
  \begin{center}
% !!!!!!!!!!!!!!!!!!!!!!!!!!!!!!!!!!!!!!!!!!!!!!!!!!!!!!!!!!!!!!!!!!!!!!!!!!!!!!!!!!!!!!!!!!!!!!!!!!!!!!!!!!!!!!!!!!!!!!!!!!!!!!!!!!!!!!!!!
    \includegraphics[width=\columnwidth]{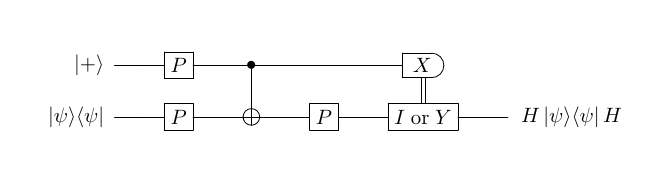}
% !!!!!!!!!!!!!!!!!!!!!!!!!!!!!!!!!!!!!!!!!!!!!!!!!!!!!!!!!!!!!!!!!!!!!!!!!!!!!!!!!!!!!!!!!!!!!!!!!!!!!!!!!!!!!!!!!!!!!!!!!!!!!!!!!!!!!!!!!
    \caption{
      The quantum circuit $\mathcal{T}$ in which the $H$ gate is applied to the single-qubit quantum state $\ket{\psi}$ (alternatively
      $\ket{\psi}\!\bra{\psi}$) with the gate teleportation technique by taking an advantage of the ancillary quantum state $\ket{+}$.
    }
% !!!!!!!!!!!!!!!!!!!!!!!!!!!!!!!!!!!!!!!!!!!!!!!!!!!!!!!!!!!!!!!!!!!!!!!!!!!!!!!!!!!!!!!!!!!!!!!!!!!!!!!!!!!!!!!!!!!!!!!!!!!!!!!!!!!!!!!!!
    \label{fig:h_gate_teleportation_calculations}
% !!!!!!!!!!!!!!!!!!!!!!!!!!!!!!!!!!!!!!!!!!!!!!!!!!!!!!!!!!!!!!!!!!!!!!!!!!!!!!!!!!!!!!!!!!!!!!!!!!!!!!!!!!!!!!!!!!!!!!!!!!!!!!!!!!!!!!!!!
  \end{center}
\end{figure}
% -----------------------------------------------------------------------------------------------------------------------------------------

Here we explicate the detailed calculations behind the teleportation of the $H$ gate. For the sake of simplicity, let us consider the
quantum circuit $\mathcal{T}$ which takes single-qubit quantum state $\ket{\psi} = \alpha\ket{0} + \beta\ket{1}$ (alternatively
$\ket{\psi}\!\bra{\psi}$) and ancillary quantum state $\ket{+}$ as an input, see Fig.~\ref{fig:h_gate_teleportation_calculations}. The
calculations before the measurement in the Fourier basis are straightforward and the two-qubit result can be written in the matrix form as
% +++++++++++++++++++++++++++++++++++++++++++++++++++++++++++++++++++++++++++++++++++++++++++++++++++++++++++++++++++++++++++++++++++++++++
\begin{align}
\ket{\phi} = \left( \mathbb{I} \otimes P \right) \circ \mathrm{C}\text{-}X \circ \left( P \otimes P \right)
\circ \left( \ket{+} \otimes \ket{\psi} \right).
\label{eq:result_before_measurement}
\end{align}
% +++++++++++++++++++++++++++++++++++++++++++++++++++++++++++++++++++++++++++++++++++++++++++++++++++++++++++++++++++++++++++++++++++++++++

Next, to implement the measurement in the Fourier basis, let us define the measurement operators as $\ket{+}\!\bra{+} \otimes \mathbb{I}$
and $\ket{-}\!\bra{-} \otimes \mathbb{I}$ for the two measurement outcomes which, by using the result in
Eq.~\ref{eq:result_before_measurement}, can be written as follows:
% +++++++++++++++++++++++++++++++++++++++++++++++++++++++++++++++++++++++++++++++++++++++++++++++++++++++++++++++++++++++++++++++++++++++++
\begin{align}
^+\!\ket{\chi} = \frac{\left(\ket{+}\!\bra{+} \otimes \mathbb{I}\right)}
                      {\sqrt{\bra{\phi} \left(\ket{+}\!\bra{+} \otimes \mathbb{I}\right) \ket{\phi}}} \ket{\phi},
\label{eq:plus_measurement_outcome} \\
^-\!\ket{\chi} = \frac{\left(\ket{-}\!\bra{-} \otimes \mathbb{I}\right)}
                      {\sqrt{\bra{\phi} \left(\ket{-}\!\bra{-} \otimes \mathbb{I}\right) \ket{\phi}}} \ket{\phi},
\label{eq:minus_measurement_outcome}
\end{align}
% +++++++++++++++++++++++++++++++++++++++++++++++++++++++++++++++++++++++++++++++++++++++++++++++++++++++++++++++++++++++++++++++++++++++++
and if we write Eqs.~\ref{eq:plus_measurement_outcome} and \ref{eq:minus_measurement_outcome} explicitly, the result will be:
% +++++++++++++++++++++++++++++++++++++++++++++++++++++++++++++++++++++++++++++++++++++++++++++++++++++++++++++++++++++++++++++++++++++++++
\begin{align}
^+\!\ket{\chi} = \ket{+} \left( \alpha\ket{-} - \beta\ket{+} \right), \label{eq:plus_measurement_outcome_explicit} \\
^-\!\ket{\chi} = \ket{-} \left( \alpha\ket{+} + \beta\ket{-} \right). \label{eq:minus_measurement_outcome_explicit}
\end{align}
% +++++++++++++++++++++++++++++++++++++++++++++++++++++++++++++++++++++++++++++++++++++++++++++++++++++++++++++++++++++++++++++++++++++++++

As one can observe, we obtain an anticipated result in case of Eq.~\ref{eq:minus_measurement_outcome_explicit}, i.e., the $H$ gate is
definitely applied to the single-qubit input quantum state $\ket{\psi}$ (alternatively $\ket{\psi}\!\bra{\psi}$). On the other hand, in
case of Eq.~\ref{eq:plus_measurement_outcome_explicit} we do not obtain an anticipated result and an additional application of the $-iY$
gate is required. The result of the $-iY$ gate application can be explicitly written as
% +++++++++++++++++++++++++++++++++++++++++++++++++++++++++++++++++++++++++++++++++++++++++++++++++++++++++++++++++++++++++++++++++++++++++
\begin{align}
\ket{+} \left( \alpha\ket{+} + \beta\ket{-} \right) = \left( \mathbb{I} \otimes -iY \right) \circ
\left( \ket{+} \left( \alpha\ket{-} - \beta\ket{+} \right) \right),
\end{align}
% +++++++++++++++++++++++++++++++++++++++++++++++++++++++++++++++++++++++++++++++++++++++++++++++++++++++++++++++++++++++++++++++++++++++++
and one can observe that we indeed obtain and anticipated result where the $H$ gate is applied to the single-qubit input quantum state
$\ket{\psi}$ (alternatively $\ket{\psi}\!\bra{\psi}$).

\end{document}